\documentclass[a4paper]{article}
\usepackage{graphicx}
\usepackage{times}
\usepackage{amssymb}
\usepackage[numbers,sort,square,comma]{natbib}
\usepackage[vcentering,dvips]{geometry}
\def\gsim { \lower .75ex \hbox{$\sim$} \llap{\raise .27ex \hbox{$>$}} }
\def\lsim { \lower .75ex \hbox{$\sim$} \llap{\raise .27ex \hbox{$<$}} }

\begin{document}

\title{Dark Matter Halos from the Inside Out}

\author{
{James E. Taylor}
\\
\\
Department of Physics and Astronomy, University of Waterloo,\\ 
200 University Avenue West, Waterloo, Ontario N2L 3G1, Canada\\
\\
Correspondence should be addressed to James Taylor, taylor@uwaterloo.ca
\\
}
\date{July 1, 2010}
\maketitle
\begin{abstract}
The balance of evidence indicates that individual galaxies and groups or clusters of galaxies are embedded in enormous distributions of cold, weakly interacting dark matter. These dark matter `halos' provide the scaffolding for all luminous structure in the universe, and their properties comprise an essential part of the current cosmological model. I review the internal properties of dark matter halos, focussing on the simple, universal trends predicted by numerical simulations of structure formation. Simulations indicate that halos should all have roughly the same spherically-averaged density profile and kinematic structure, and predict simple distributions of shape, formation history and substructure in density and kinematics, over an enormous range of halo mass and for all common variants of the concordance cosmology. I describe observational progress towards testing these predictions by measuring masses, shapes, profiles and substructure in real halos, using baryonic tracers or gravitational lensing. An important property of simulated halos (possibly {\it the} most important property) is their dynamical `age', or degree of internal relaxation. The age of a halo may have almost as much effect as its mass in determining the state of its baryonic contents, so halo ages are also worth trying to measure observationally. I review recent gravitational lensing studies of galaxy clusters which should measure substructure and relaxation in a large sample of individual cluster halos, producing quantitative measures of age that are well-matched to theoretical predictions. The age distributions inferred from these studies will lead to second-generation tests of the cosmological model, as well as an improved understanding of cluster assembly and the evolution of galaxies within clusters.
\end{abstract}

\section{Introduction} 
\label{sec:intro}
Everywhere in the universe, on scales comparable to the size of galaxies or larger, the effects of gravity appear to be anomalously strong. The earliest of these observations dates to 1933, in work by Zwicky on the Coma cluster of galaxies \citep{Zwicky}. Using redshifts derived from optical spectroscopy, Zwicky  measured a velocity dispersion of 1600 km/s for the galaxies in Coma, indicating that the cluster had an enormous amount of internal kinetic energy. For the structure to be dynamically stable, the gravitational potential energy required was 20 times what would have been inferred from the distribution of visible stars and gas alone. Zwicky suggested that some sort of `dark matter', cold enough to be effectively invisible in his optical observations, might account for the deficit. The term `dark matter' has been with us since, although the idea remained dormant for nearly 40 years following Zwicky's discovery. 
A generation later, the idea of dark matter was rediscovered by Rubin and Ford. Starting with a study of the Andromeda galaxy, M31 \citep{Rubin}, they gradually established that rotation speeds in the 
outskirts of spiral galaxies were far higher than would be 
possible for a stable system bound together by its visible gas and stars alone. Once again, the mass needed to explain the observations was enormous, and the concept of a `dark halo' around the galaxy was introduced. It was unclear at the time whether this matter was normal, baryonic\footnote{That is matter whose mass was comprised of baryons -- protons and neutrons.} matter in the form of cold gas or compact objects, or whether it was some novel form of non-baryonic matter.

Although individual galaxies like M31 or clusters of galaxies like the Coma cluster are still powerful probes of the nature and distribution of dark matter, the evidence for dark matter (and/or non-baryonic matter) is now much more broadly based than it was in the 1930s or the 1970s. The spectrum of fluctuations in the cosmic microwave background (CMB), for instance, strongly constrains the composition of the universe when it was less than a million years old. At this time the radiation density is high enough that interactions with photons have a substantial effect in smoothing out variations in the density of normal matter (that is matter with an electromagnetic coupling). Weakly-interacting matter is free to go about its business, however, and fluctuations in this component can grow once the corresponding particle has cooled sufficiently to become non-relativistic. The relative heights of the second and third acoustic peaks in the CMB angular power spectrum constrain the ratio of the two components; together with the first peak they indicate that the total matter density in the universe, as a fraction of the critical density, is $\Omega_m = 0.267 \pm 0.029$, while the fraction in baryonic matter coupling to photons is only  $\Omega_{bar} = 0.045 \pm 0.003$ 
\citep{Larson} . The difference between the two figures must come from a component with interactions of the weak scale or less. In principle this dark component could have decayed since that time, but measurements of large-scale structure (e.g.~\citep{Sloan}) and cosmic expansion (e.g.~\cite{SNLS}) find a consistent value for the total matter density at low redshift. Meanwhile, abundances of light elements provide an independent estimate of the baryon density that is roughly consistent with CMB measurements and much lower than the density needed to explain large-scale structure or cosmic expansion \citep{Cyburt}. Thus the best evidence for cold dark matter -- or more specifically for a non-baryonic, pressureless component that dominates the matter density and is non-relativistic at early times -- now comes from  the largest scales in the universe. This point is sometimes missed by alternative theories that focus exclusively on galaxy rotation curves or similar tests. 
 
The idea of a weakly-interacting particle massive enough to be `cold' (or non-relativistic) at the time of the CMB is not unwelcome in proposed extensions to the Standard Model of particle physics. To explain the many cases of fine tuning or the strange discrepancies in scale in the Standard Model, these extensions introduce new high-energy symmetries and new families of massive particles, partnered to the familiar low-energy states. Supersymmetry, with its family of massive super-partners, is the most commonly cited example, but the Kaluza-Klein modes from higher-dimensional theories are similar in many respects, and other well-motivated models exist \citep{Feng}. There is great hope that the Large Hadron Collider will produce one of these candidates in the near future, that  direct detection experiments will detect it in the lab, or that indirect detection experiments will detect its annihilation or decay products.  Not all candidates are easily detectable, however; some may elude most or all attempts to identify them \citep{Hooper}. Furthermore, determining the full properties of a candidate will probably require astrophysical input as well as laboratory or accelerator measurements \citep{Hooper}. Thus astrophysics will remain an important source of information about dark matter for some time to come. The cosmic abundance, mass, decay channels, self-couplings or couplings to known particles and possibly even excited states of the dark matter particle(s) may all eventually be derived from astrophysical measurements \cite{Bertone}.

As a central component in the current picture of cosmological structure formation, cold dark matter (CDM) has been extremely successful. The standard cosmological model of structure formation (which I will refer to loosely as the `CDM model' although it contains many other ingredients) posits that the universe contains known particles -- baryons, photons, a small contribution from warm or hot neutrinos -- but also two dominant dark components, weakly-interacting cold dark matter and a cosmological constant or some similar form of  `dark energy'. Given these ingredients and starting from an inflationary power spectrum at early times, the CDM model predicts the subsequent growth of fluctuations in the matter distribution, the 3-D power spectrum of these fluctuations after radiation-matter equality, the angular power spectrum of temperature fluctuations in the CMB at the time of last scattering, and the properties of large scale structure post-CMB. The predictions are in such good agreement with large-scale measurements of these quantities that it is hard to construct sensible alternatives to the standard picture. 

On somewhat smaller scales, the CDM model does not specify {\it how} galaxies form or evolve, but it does suggest where they may form. Analytic theory and numerical simulations of structure formation show how fluctuations grow into virialized halos and predict the abundance and clustering of dark matter halos as a function of mass and redshift. Empirical models can then be used to place galaxies in these halos in a way that is consistent with the galaxy abundances and clustering measured in surveys. Finally, measurements of the average gravitational potential associated with galaxies, e.g. through gravitational lensing, close the loop and test the consistency of whole model. Here too, the abundance and clustering pattern of galaxies and their association with excess gravitational potential are now thoroughly established, so much so that any alternative model of structure formation is constrained to make predictions very similar to the standard CDM theory at the redshifts we have observed to-date. 

In short, the CDM model has passed all observational tests on scales of individual halos or larger. In the process, the observational evidence has narrowed the field of competition for alternative models
considerably. {\it Any} working theory of structure formation, whatever its physical basis, is now constrained by observations to look a lot like the CDM model. An unfortunate corollary is that 
large-scale tests of structure formation may no longer be enough to make progress in this field. Future observations of large-scale structure, e.g. from forthcoming experiments to measure baryon acoustic oscillations \cite{Wiggelz, BOSS, BigBOSS}, may teach us something about the nature of dark energy,  but they will probably not teach us much about the nature of dark matter. For this, we will need more detailed studies of the individual halos detected by these spectroscopic surveys and by the next generation of wide-field imaging surveys \cite{Hypersuprime, DES_URL, PANSTARRS_URL, LSST_URL}.  

Indeed, while models of structure formation are highly constrained by large-scale observations, the constraints on small scales are considerably looser. If the standard CDM model is correct, luminous matter does not trace most dark matter structure below the scale of bright galaxies (halos of mass $10^{11}$--$10^{12} M_\odot$), and does not trace {\it any} of it below the scale of the smallest dwarf galaxies (halo masses of $10^6$--$10^7 M_\odot$). In the default model, the dark matter candidate is massive, cold, weakly interacting and stable. In this case, dark matter structure extends down to scales of 0.1 A.U. or less (halo masses of $10^{-6} M_\odot$ or less) and is almost scale-invariant all the way down \citep{Diemandsmall}. To test this incredible prediction of 10 decades or more of invisible structure filling our universe, we need to study dark matter in the highly non-linear regime, deep in the heart of halos. This is where the smallest and oldest dark matter structures end up, and it is also where dark matter reaches its highest density and where new physics -- scattering, annihilation or decay into other particles -- should be most evident. In fact, since the Milky Way is embedded in a dark matter halo and we reside relatively close to its centre (within the central 3\% in radius), any local study of dark matter must come to grips with the highly non-linear regime of CDM structure formation. In that sense, our view of dark matter halos is necessarily an `insider's view'.

In this article I review some of what we know and what we can learn about the internal structure of dark matter halos. The literature in this field is extensive, so rather than provide an exhaustive survey I have focussed on a few key concepts and simple results. Since reviews on this subject are unfortunately rare, I have tried to include enough basic explanatory material to make the main results accessible to a broader audience of non-specialists. In section \ref{sec:nonlinear}, I first introduce some basic elements from the standard theory of structure formation, and clarify why halos correspond to `non-linear' structure. Section \ref{sec:universal} reviews the surprising universality of halo properties that has emerged from numerical simulations, explains how halos grow and evolve through mergers and accretion, and introduces the concept of halo `age' as a description of the degree of internal relaxation. Section \ref{sec:obs} considers methods for determining halo ages observationally, using semi-analytic models of halo substructure to evaluate their effectiveness. Finally in section \ref{sec:discuss} I discuss the prospects for these methods in current and future observations. I do not discuss specific dark matter candidates or their properties in detail in this article, since these have been extensively reviewed elsewhere (e.g.~\cite{Bertone} and articles in this volume), but I summarize in section \ref{sec:discuss} how measurements of halo properties can help constrain these candidates and other aspects of fundamental physics.

\section{From Linear to Non-linear Structure Formation} 
\label{sec:nonlinear}

It is worth clarifying how dark matter halos and non-linear structure relate to the linear physics of fluctuations in the early universe. Reviews of basic cosmology and the linear perturbation theory can be found in almost any textbook on cosmology (e.g. \cite{Ryden} for an elementary introduction, or \cite{Peacock} or \cite{Padmanabhan} for more advanced treatments). A more specific discussion of large scale structure can be found in \cite{Peebles}. An excellent review of Press-Schechter theory is given in 
\cite{Zentner08}.

If we perturb the smooth matter distribution in a region of the early universe by a small amount 
$\delta = (\rho - \bar{\rho})/{\bar\rho}$ around the average density $\bar{\rho}$, then the perturbation 
will obey:
\begin{equation}
{\ddot{\delta}} + 2H{\dot{\delta}} - {3\over 2}\Omega_mH^2\delta = 0,
\end{equation}   
where $a = 1/(1+z)$ is the scale factor, $H = \dot{a}/a$ is the Hubble constant and $\Omega_m = \bar{\rho}/\rho_c$ is the matter density relative to the critical density $\rho_{c}$.
Solving this perturbation equation in a flat, matter-dominated universe with $\Omega_m = 1$ and $H = 2/(3t)$, we find solutions that grow as $D(t) \propto t^{2/3}$ and solutions that decay as $D(t) \propto t^{-1}$. Since $a \propto t^{2/3}$, this means that the growing mode will increase in amplitude linearly with scale factor and independent of amplitude. For more general cosmologies the relative amount of growth is no longer proportional to $a$ (see \cite{Hamilton} for a general discussion), but it will still be independent of amplitude $\delta$ provided the amplitude is small. If we decompose a given pattern of fluctuations into distinct Fourier modes $\delta_{\mathbf k}$ of (spatial) wavevector ${\mathbf k}$, linear growth will preserve the relative phases and amplitudes of the different modes since it is independent of $\delta$. Thus the spatial pattern of fluctuations and the shape of the corresponding power spectrum will be preserved as long as the amplitude of fluctuations remains small. This feature of linear growth makes it particularly easy to compare fluctuations at different epochs and has led to the highly developed statistical machinery used to study the CMB and large-scale structure. 

At early times, the matter density field resulting from inflation should consist of Gaussian fluctuations with uncorrelated phases around a nearly critical mean density. Since the phases and relative amplitudes of fluctuations in this field are unchanged during subsequent linear growth, we can describe them independent of redshift by dividing their amplitude by a linear growth factor $D(z)$ that measures the amount by which linear growth has amplified fluctuations by redshift $z$ relative to some reference epoch\footnote{Note that the linear growth factor is sometimes defined as $g = D(a)/a$, and $D(a)$ referred to as the amplitude of the growing mode, e.g. in \cite{Hamilton}.}. Normalizing the growth factor to unity at $z = 0$, this is equivalent to considering the initial field as it would be if evolved linearly to the present day. Smoothing this linearly-evolved field on a scale $R$ we obtain a new Gaussian random field of variance $\sigma^2(R)$, and this variance alone is enough to characterize the matter distribution statistically on these scales. The function $\sigma(R)$ summarizes all information about the (linear) power spectrum, while all information about linear growth is contained in $D(z)$.

As fluctuations grow to amplitudes $\delta \sim 1$, it is clear their growth must accelerate beyond the linear rate. Since the early universe is very close to the critical density, positive density fluctuations exceed this limit and represent regions with net positive curvature. These regions will eventually stop their expansion, turn around and recollapse, at which point their density diverges formally. In 1972, Gunn and Gott  considered the behavior of a spherical region of uniform (over-)density and showed that it would recollapse at a well-defined time \citep{Gunngott}. The collapse time can be expressed in terms of the initial conditions at some early time when the amplitude of the fluctuation is small; it corresponds to the time by which the initial fluctuation, had it grown at the linear rate, would have reached a critical threshold $\delta_c$. This critical threshold has the value $\delta_c  = 3/5(3\pi/2)^{2/3} \sim 1.69$ for a wide range of cosmologies (e.g.~\cite{Peebles, Laceycole}). 

This simple result leads to a very clever analytic estimate of the abundance of collapsed halos, derived by Press and Schechter \cite{Pressschechter}. Consider a region which has a spherically-averaged density contrast $0 < \delta(z) < \delta_c$ at redshift $z$  (or scale factor $a = 1/(1+z)$). The region will recollapse by the present day if its linearly-evolved density contrast exceeds $\delta_c$ at $z = 0$. We can write this condition:
$$\delta(0) = \delta(z)/D(z) > \delta_c,$$
where $D(z)$ is the linear growth factor normalized to unity at the present day and $\delta(0)$ is the amplitude of the fluctuation evolved linearly to the present day. 
The fraction of the linearly-evolved matter density field $F$ contained in collapsed objects at redshift $z = 0$ is thus simply the fraction of regions with density contrast $\delta(0) > \delta_c$, that is:
$$F(0,R) = {{1}\over{\sqrt{\pi}}} \int_{\delta_c}^\infty \exp \left({{-\delta^2} \over{2\sigma^2(R)}}\right) 
= {1\over 2}{\rm erfc}\left(\nu(0)/\sqrt{2}\right)\ ,$$\label{eq:psfrac}
while the fraction contained in collapsed objects at a higher redshift z is:
$$F(z,R) = {{1}\over{\sqrt{\pi}}} \int_{\delta_c/D(z)}^\infty \exp \left({{-\delta^2} \over{2\sigma^2(R)}}\right) 
= {1\over 2}{\rm erfc}\left(\nu/\sqrt{2}\right)\ ,$$\label{eq:psfrac2}
where  $\nu(0) =  \delta(0)/\sigma(R)$ and $\nu =  \delta(z)/\sigma(R) = \delta_c /\left(D(z)\sigma(R)\right)$  are the normal variates of the smoothed Gaussian random field and erfc is the complementary error function. The two integrals differ only in their lower limit; since less growth has occurred by redshift $z$ a smaller faction of the field will have collapsed, so the threshold for collapse must be higher. 

The insight of Press and Schechter was that if a fraction $F(R)$ of the matter distribution met the collapse criterion when smoothed on a scale $R(M) = (3M/4\pi\bar{\rho})^{1/3}$, and the fraction was reduced to $F' < F$ when the smoothing scale increased to $R' = R(M + dM)$, then the difference could be assumed to have collapsed to form objects in the mass range $[M, M+dM]$. Thus differentiating the previous equation 
with respect to filtering scale (or equivalently enclosed mass) leads to the Press-Schechter expression for the halo mass function:
$${{dn}\over{dM}}\, dM = 2{{\bar{\rho}}\over{M}} {{dF}\over{dM}}(z, R(M))\, dM = {{\bar{\rho}}\over{M}}\sqrt{2\over{\pi}} \exp\left({{-\delta^2} \over{2\sigma^2(R)}}\right) {{d\nu}\over{dM}}\, dM\ ,$$
\label{eq:psmf}
where the factor $\bar{\rho}/M$ converts from mass density to number density. An extra factor of two has crept in here to correct for the under-counting of underdense regions -- a subtle point in Press-Schechter theory (see \cite{Bond} for a rigorous derviation). 

The beauty of the Press-Schechter approach is that it relates the abundance of non-linear halos to fundamental quantities from linear physics -- the growth factor $D(z)$ and the spectrum of fluctuations, represented by $\sigma(M)$. The model can also be extended to calculate conditional probabilities, e.g.~the probability that a point will be contained in a collapsed region of mass $M_1$ at redshift $z_1$ and then in a region of mass $M_2$ at redshift $z_2$. I will discuss extended Press-Schechter (EPS) theory and these conditional statistics further in section \ref{sec:obs}. In its simplest form, however, Press-Schechter theory has many inconsistencies and only approximately describes the behavior seen in simulations (e.g.~\cite{Robertson}). 
I will discuss more recent numerical work on halo evolution in the next section.

What do analytic arguments tell us about the internal structure of dark matter halos? The collapse of a completely cold spherical shell, e.g.~in the model of Gunn \& Gott discussed previously, would produce a spike of infinite density at the centre of the perturbation. Real perturbations will only reach finite density, however, because even cold dark matter has some residual random motion relative to the Hubble expansion, and because the region surrounding a perturbation will not be perfectly spherically symmetric. Both these effects add angular momentum to the orbits of infalling particles, keeping them away from the point at $r = 0$. In general, any deviation from symmetry will be amplified by the collapse, leading to a mixing of orbits and the rapid `virialization' of the system, through which its central region reaches the virial equilibrium:
$$W = -2K$$
between potential energy $W$ and kinetic energy $K$. One of the main results of the spherical collapse model of Gunn \& Gott is an estimate of the final density a virialized region achieves. This can be derived from energy conservation, which shows that the final size of a spherical region after collapse will be half its size at `turn-around' (the moment at which its radial velocity is instantaneously zero). Meanwhile the surrounding universe will continue to expand, increasing the contrast between the collapsed region and the mean density of the universe. A numerical calculation (e.g.~\cite{Padmanabhan}) shows that the density contrast relative to the background density $\bar{\rho}$ will be $\Delta_c \equiv \rho_{vir }/\bar{\rho} = 18\pi^2 \sim 178$ in an Einstein-deSitter universe, or a slightly larger value in LCDM cosmologies.

Having reached this equilibrium, the virialized region no longer feels the universal expansion around it and in the absence of subsequent accretion, its physical size will remain constant with time. This idealized situation is never achieved for any real halo, however. Since a halo represents a region with $\delta > 0$, one can always find a larger region around it with smaller density contrast $\delta'$ such that $\delta > \delta' > 0$. Since this larger region also has net positive curvature, it will recollapse in turn, adding new material to the virialized halo within it. Thus halos never remain isolated, but continue to accrete material from the universe around them. While their central regions are in approximate equilibrium, some new material is being mixed into these regions at all times, and their outer regions are constantly growing to include more and more mass. In a system like the Milky Way, for instance, the virial radius (that is the radius within which $W \sim -2K$) is estimated to be around 300 kpc (\cite{Watkins, Limass} and references therein), but matter is accreting onto the halo from a much larger region extending out to 
$\sim 1$Mpc. Since virialization produces a constant density contrast relative to the background, all halos at a 
given redshift are predicted to have the same mean density interior to their virial radius, and this virial density decreases with time as the background density decreases. 
 
Averaging over large scales, the density field of the universe can be decomposed into several distinct components (e.g.~\cite{Hayashixcor}): 1) the linear regime of large-scale, low-amplitude fluctuations with fixed comoving size and a linearly growing amplitude; 2) the collapsed, virialized regions corresponding to individual halos, whose sizes and densities evolve slowly through mergers and accretion; and 3) a quasi-linear regime which interpolates between these two. The division into different regimes is made formal in ``halo models", which decompose the clustering of galaxies into clustering within a single halo on small scales (the ``one-halo term") and clustering of multiple halos due to linear or quasi-linear fluctuations in the large-scale matter distribution (the ``two-halo" term) \cite{Cooray}. In the present-day universe, the division between the two regimes occurs on scales of roughly 1 Mpc. In what follows I will consider the matter distribution within halos, which determines the non-linear, or ``one-halo", contribution to the matter distribution.

\section{The Universality of Halo Properties}
\label{sec:universal}

\subsection{A Universal Density Profile}
Neither the spherical collapse model nor Press-Schechter theory alone specify what the internal structure of dark matter halos should be. One analytic approach to estimating this is to generalize the spherical collapse model, considering the one-dimensional collapse of concentric shells of different radii and densities. A sensible choice for the initial run of density with radius in such a model is the matter (auto-)correlation function, which describes clustering statistically in terms of the average excess density of matter around a point, and can be derived from the power spectrum. Several early analytic models of this kind, notably \cite{Gunngott, Goldreich, Bertschinger}, predicted that matter would cluster around a point to produce a radial density profile that was a steep power-law with a constant slope.

Despite these analytic insights, real progress on the matter distribution inside halos did not occur until numerical simulations of structure formation started to resolve individual halos. Building on initial work which made it clear that halo density profiles were not simple power laws \cite{Quinn,  Frenk, Dubinski, Crone}, Navarro, Frenk and White \citep{NFW96, NFW97} determined that the halo density profile was well-fit by a single  functional form:
$$\rho(r) = { {\rho_s r_s^3} \over {r(r_s + r)^2} }.$$
Not only was this form not a simple power law; with suitable choices of $r_s$ and $\rho_s$ it seemed to fit {\it all} the simulation results, independent of halo mass, power spectrum and cosmology. The profile, since named the Navarro-Frenk-White (NFW) profile or universal density profile (UDP), was the first of many indications of the surprising simplicity of halo properties.

The radial profiles of halos have a number of interesting features. First, rather than being scale invariant they contain a characteristic length $r_s$, the NFW {\it scale radius}, at which the logarithmic slope of the profile is ${\rm d}\ln\rho/{\rm d}\ln r = -2$. This scale is often defined with respect to the virial radius $r_{vir}$ via the{\it \ concentration parameter}  $c \equiv r_{vir}/r_s$. As discussed below, the scale radius seems to mark the division between two phases in the assembly of the halo, a rapid phase where the central part of the profile builds up with  $\rho(r) \propto r^{-1}$, and a slower phase where an outer envelope forms around the halo with a steeper profile going as $\rho(r) \propto r^{-3}$. The velocity distribution inside halos is roughly isotropic, that is $\sigma_{v,r} \sim \sigma_{v, \theta} \sim \sigma_{v,\phi}$, although the outer regions have a slight radial bias, probably reflecting the continued infall of material on radial orbits \cite{Navarronew}. Perhaps the greatest surprise is the ``pseudo-phase space density" $\rho(r)/\sigma^3(r)$ one can construct from the density and velocity dispersion profiles. In the relaxed part of the halo this quantity is a featureless power-law, suggesting the possibility of a fundamental connection between the inner and outer parts of the density profile \cite{TN01}.
  
\subsection{Universal Patterns in Halo Growth}

As suggested above, the features of the UDP seem to be connected to the evolutionary history of halos. The growth of an individual halo can be described by specifying its total mass (say the mass within the virial radius, defined as the region with mean density $\Delta_c$ times the critical density) as a function of redshift or scale factor. In what follows I will refer to this function $M(z)$ as the ``mass accretion history" (MAH) of a halo and often consider it normalized by the value $M(z=0)$. Studies of halo growth in N-body simulations show that MAHs have a characteristic shape, consisting of rapid growth at early times and slower growth at late times, with the break point between the two varying from halo to halo. They have be fitted to the functional form $M(z)/M(0) = (1+z)^\beta$ \cite{vandB} as well as to the form $M(z)/M(0) = \exp(-\alpha z)$ \cite{Wechsler}, but more recent work \cite{Tasitsiomi, Mcbride} makes it clear both terms are required to fit the full range of MAHs seen in simulations. In \cite{Mcbride}, McBride et al.~fit a general 2-parameter form for the MAH: 
$$M(z)/M(0) = (1+z)^\beta\exp(-\gamma z)\, ,$$ 
but find that the fitted values of $\beta$ and $\gamma$ they derive are strongly correlated in their ensemble of MAHs, suggesting an even better parameterization may exist.

These analytic approximations to the typical MAH are particularly intersting because of the strong connection between the MAH of a halo and its density profile. Crudely speaking, when halos grow rapidly their density profiles stay relatively shallow. From the work of \cite{Zhao03}, the scale radius $r_s$ appears to increase in sync with the virial radius during these growth spurts, such that the concentration parameter stays at a roughly constant value $c \simeq 4$. During slower phases of evolution, while the virial radius continues to grow roughly as the scale factor of the universe (since the virial density contrast is approximately constant), the scale radius stays constant. Thus concentration increases as scale factor during these quiescent periods. Based on these patterns, Wechsler et  al.~\cite{Wechsler} and Zhao et al.~\cite{Zhao03} provide similar algorithms for predicting a halo's concentration at any time, given its MAH. (See also Lu et al. \citep{Lu06}, which ties the precise shape of the density profile to the MAH.) In the particularly simple model of Wechsler et al., $c = c_0 (a/a_c)$, where $c_0 \sim 4$ is the concentration of a halo undergoing rapid accretion and the term $a/a_c$ is the ratio of the present scale factor $a$ to the scale factor $a_c$ at the time the halo last stopped accreting rapidly. Present-day galaxy halos have typical concentrations of $\sim$12, indicating that they stopped growing rapidly at $a_c = 1/3$ or $z = 2$, while galaxy clusters have concentrations of 4--6, indicating that they have just formed recently ($a_c = $1--2/3, or $z = $0--0.5).

\subsection{Halo Shape and Spin}

Just as simulated halos have a well-defined distribution of MAHs, correlated with their concentration parameter, so too they show a regular and universal distribution of shapes. Halos are generally triaxial, with axis ratios of $b/a \sim c/b \sim 0.6$--0.8 (e.g.~\cite{JingSuto, Allgood, Bett}) (although their potentials may have slightly different shapes -- cf.~\cite{Hayashishape}). Typical examples are more often prolate (cigar-shaped) than oblate (disk shaped). Shape is correlated with age or merger history (e.g.~\cite{Bett}), although this correlation has been established only in an average sense. A physical mechanism which may account for the correlation was outlined in \cite{Mooremerg}. A system of two halos merging on a radial orbit can be described in terms of a tensorial version of the virial theorem, in which the contributions of kinetic and potential energy must eventually reach an equilibrium component by component. Dissipationless mergers between dark matter halos should roughly conserve the individual tensor components, such that the final merger remnant will remember the original orientation of the infalling pair and will remain more extended along that axis. From this point of view, a halo's shape may provide  an interesting clue to the orientation of the merger that formed it. Tests of this idea are possible using gravitational lensing (e.g.~\cite{Mandelbaum}) or the orbits of satellite galaxies (e.g.~\cite{Brainerd}) to measure halo ellipticity with respect to local structure. 

The internal velocity distribution of halos also shows regular patterns. In particular, halos seem to have a fairly universal distribution of internal angular momentum \cite{Bullock01} and of net spin (\cite{Bett} and earlier references therein). Overall the spin is small; defining a dimensionless  spin parameter $\lambda$ from the size, mass and energy of a system, the typical value for halos is $\lambda = 0.03$--0.05. Thus spin contributes little to the support of the system against gravity. The angular momentum distribution within halos presumably relates to the MAH and to the features of the density profile (\cite{Bett} show that rounder halos have less net spin on average, for instance), although the connection is still somewhat unclear.

\subsection{Halo Substructure}

The standard picture of halo formation suggests halos should contain dense substructure, corresponding to the visible baryonic structure in groups and clusters.
Halos accrete matter continuously from their surroundings, and this matter may include other virialized halos. In fact, in CDM cosmologies the shape of the power spectrum is such that the variance $\sigma^2(M)$ increases at small masses, so small regions have a greater range of density contrast than large ones and are more likely to cross the threshold for collapse at early times. As a result, at any time there are many more small halos in the universe than large ones. The smallest halos collapse and virialize at the earliest times on average, reaching high virial densities. They may then be incorporated into larger halos where their cores can survive as dense substructures, or ``subhalos". The most massive subhalos (i.e.~those closest to their parent halo in mass) evolve quickly, but otherwise the evolution is relatively independent of subhalo mass. Thus the mass spectrum of subhalos within a halo at any given time reflects the average cosmic mass function of halos present in the accreted material, although the normalization of the subhalo mass function decreases slowly as individual subhalos lose mass through tidal heating and stripping. 

From this picture, substructure should correlate with the `age' of the halo, in the sense of the mean time since its mass was assembled into a single object. For a given initial spectrum of subhalos, this prediction can be made quantitative by determining the mean number of orbits subhalos have 
spent in the parent halo and assuming a certain amount of mass loss occurs once per orbit at pericentric passage. Slightly more elaborate analytic or semi-analytic models of the evolution of the subhalo mass function were developed in \cite{TB01, TB04, vandBtime}. Beyond the systematic variation with halo age, the shape of the subhalo mass function is approximately universal. It consists of a power-law with an exponential cutoff at $M_{sat}/M_{main} \sim 0.1$, as discussed below (see also \cite{Gaorecent, Giocoli}).

While the relationship between simulated halos and observed field galaxies seems reasonably straightforward, the connection between subhalos and group or cluster members has 
historically been much harder to understand. The earliest simulations found no substructure 
at all --  the ``overmerging" problem, which turned out to be due to a lack of mass and force 
resolution 
\cite{Moorerev}.
After finally resolving halo substructure in clusters \cite{Tormen, Ghigna, Klypinover}, simulations quickly started producing too much of it in group or galaxy-sized halos \cite{Klypin99, Moore99} -- far more than was needed to host visible dwarf galaxies in the Local Group, for instance. This issue continues to generate controversy, but is beyond the scope of this article (see \cite{Kravtsovrev} for a recent review). Suffice it to say that if the LCDM model is correct, the current generation of high-resolution simulations of individual halos \cite{Aquarius, Vialactea} indicate the existence of a huge amount of substructure around the Milky Way that is not traced by baryons.
 
\subsection{The Dynamical Evolution of Halos}

So far I have focussed on the static properties of halos, but halos are dynamical systems. As mentioned above, they are constantly accreting new material, both relatively smooth matter and a spectrum of other halos. Mergers of one halo with another have long interested simulators, as they may give rise to galaxy mergers and drive some of the more spectacular evolution in galaxy morphology. Thus there has been extensive analytic theory and numerical work on this subject. I will summarize some salient points here. 

In the spherically-symmetric limit of halo growth, subhalo orbits are expected to be purely radial and start their infall with a fixed energy corresponding to their potential energy at turn-around. In real halos, departures from symmetry in the immediate surroundings of a system will scatter the initial energy and angular momentum of satellites. The angular momentum distribution of subhalo orbits at infall (and to a lesser extent the energy distribution) have been studied in detail by many authors (\cite {Tormen, Ghigna, Kochfar, Benson}); see also recent work by \cite{Kang, Wetzel}. The angular momentum of an orbit can be characterized by its circularity parameter $\varepsilon = L/L_c$, where $L$ is the angular momentum of the orbit and $L_c$ is the angular momentum of a circular orbit with the same energy. In terms of $\varepsilon$, the initial orbital 
distribution is almost uniform, with a slight peak around $\varepsilon = 0.5$. Given the steep density profile of dark matter halos, this angular momentum distribution produces mainly radial orbits, with a mean axis ratio of 6:1 \cite{Ghigna, vandBorbits}.

Subsequent to the merger, subhalo properties evolve under the influence of dynamical friction, 
tidal mass loss and tidal heating. Dynamical friction drags in the most massive satellites, but becomes negligible below $M_{sat}/M_{main} \sim 0.001 ($\cite{Colpi, Taffoni,  TB04}). Tidal mass loss peaks strongly once per orbit at passage through pericentre. This episodic mass loss should produce coherent streams of (dark matter) debris like those seen in disrupting stellar systems. After detaching from the progenitor subhalo, tidal streams evolve quite simply with time, and their physics is well understood (e.g~\cite{Johnston, Helmi}). Because mass loss, heating and stream formation occur once per orbit at the time of pericentric passage,
the characteristic timescale for evolution in halo properties is the timescale for radial oscillations in the orbit, $P_{rad} = 2\pi/\kappa$, where $\kappa$ is the epicyclic frequency. At any redshift, this timescale is comparable to the  instantaneous Hubble time $H(z)^{-1}$ \citep{TB04}. Beyond the first few orbits, the longer-term evolution of subhalos is an unresolved problem. It is not clear when (if ever) subhalos are disrupted by repeated mass-loss \cite{Hayashi}. This issue is particularly relevant to the fate of the smallest subhalos, which may or may not have survived for hundreds of orbits in the halo of the Milky Way \cite{Diemandsmall, Zhao, Greendis, Berez}.

{\vskip 0.5cm}

In summary, halos show many simple trends and properties that are universal, in the sense that they apply independent of mass or cosmology. The universal density profile is the best known of these properties, but the regular trends in mass accretion history (MAH) and the correlations of other quantities with the MAH deserve further consideration. They suggest that age is possible the most important property of a halo. Recently assembled halos have massive substructure, non-spherical shapes, low concentration parameters and possibly more spin and/or velocity anisotropy. Older halos are generally smoother, more spherical and more concentrated. In next section I consider one specific example of a trend with age, the correlation between age and substructure, in more detail.

\section{Defining and Measuring Halo Age}
\label{sec:obs}

If age is the organizing principle in halo properties, there remains the question of how to define and measure it. A pragmatic approach is simply to construct model halos, study them with mock observations and see what observations produce the clearest determination of age or formation history. A large number of model halos are required to fully explore the multivariate distributions in halo properties, so semi-analytic halo models are a convenient tool to use for this task. In this section I will use the model developed in \cite{TB01, TB04, TB05a, TB05b} to study halo substructure and ask how well lensing or X-ray studies of cluster substructure can determine the cluster age distribution.

\subsection{Models}

The semi-analytic model introduced in \cite{TB01} and \cite{TB04} consists of two components: a Monte-Carlo algorithm for generating random merger histories, or merger `trees', for individual halos, and an 
analytic description of subhalo evolution which is applied to each satellite subhalo as it 
merges with the main system. I summarize each component briefly below; for a full description see
\cite{TB01, TB04, TB05a, TB05b}.

\subsubsection{Merger Trees}

In the extension to Press-Schechter mentioned in Section \ref{sec:nonlinear}, the growth history of a halo is determined by the distribution of mass around it, averaged spherically on successively lager scales. Ignoring higher-order correlations, this distribution is Gaussian with variance $\sigma^2(M)$ when averaging on a scale  $R = (3M/4\pi\bar{\rho})^{1/3}$, where $\bar{\rho}$ is the mean density of the universe. The entire history of a halo can be generated by taking random walks in density fluctuation $\delta$,  using a Gaussian normal variate scaled by $\sigma(M(R))$, as the scale $R$ decreases from infinity (where $\delta = 0$, since the density must equal the mean value by definition) to zero, where $\delta$ may diverge. The resulting trajectory $\delta(R)$ or $\delta(M)$ is then mapped onto evolution with time or redshift using the condition for collapse $\delta > \delta_c(z) = \delta_c(0)/D(z)$. Since $\delta$ can increase or decrease with scale, the trajectory must be filtered such that $\delta(R)$ increases monotonically as R decreases; this corresponds to finding the {\it largest} value of $R$ in the trajectory 
with a given $\delta$ and assuming that that scale collapses when $\delta > \delta_c(z)$ \cite{Laceycole}.

This process produces a single, randomly-generated but representative mass accretion history $M(z)$. Because of the filtering operation, this trajectory may have discontinuous jumps in mass. These are interpreted as mergers, in which the mass of the halo instantaneously increases by a finite amount.
If each merger involves a single other halo, we can associate the mass change with this new halo and follow its evolution along with the main branch. By following every merged halo in every branch iteratively down to some mass resolution limit, we can generate a `merger tree' describing the entire formation history of our original object.

There are many subtleties to generating accurate merger trees, and no one method is ideal. A recent comparison of methods is given in \cite{Macomparison}. An analytic expression for the merger probability  in a given redshift step is available only for binary mergers in which two halos merge, but these events do not normally account for the whole mass of the final system, leading to additional branches in the merger tree and/or accreted mass below the mass resolution of the tree. The trees used in this section were generated using the method of Somerville \& Kolatt \cite{SK}, which picks time steps short enough that binary mergers are very likely in the tree. In practice, this method breaks the main progenitor down into small progenitors or accreted mass slightly too quickly, leading to mass accretion histories that are slightly too young; this is a common problem for many merger tree algorithms \cite{vandB, Cole, Macomparison}. 

Part of the inaccuracy may come from the conditional probability used in the method. In standard EPS theory \cite{Laceycole}, the probability of going from a halo of mass $M_1$ at redshift $z_1$ to a halo of mass $M_2$ at redshift $z_2$ can be written as a simple function of a single variable
$\Delta\nu = \Delta(\delta_c(z))/\sqrt{\Delta(\sigma^2(M))}$, where $\Delta(\delta_c(z)) = \delta_c(z_1) - \delta_c(z_2) = \delta_{c,0}(1/D(z_1) - 1/D(z_2))$ and $\Delta(\sigma^2(M)) = \sigma^2(M_1) - \sigma^2(M_2)$. This function is simply the unconditional probability from the mass function, shifted to a new origin in $\nu$:
$$P(\Delta\nu)dM = P_{PS}(\Delta\nu)dM = \sqrt{2\over \pi} \exp\left({-\Delta\nu^2}\over 2\right) {{d(\Delta\nu)}\over{dM}}dM\, .$$

This conditional probability assumes a collapse threshold independent of mass: $\delta_c(M,z) = \delta_c(z)$, such that $\Delta(\delta_c) = \delta_c(z_1) - \delta_c(z_2)$. This would be the case for spherical halos; for non-spherical halos the collapse 
threshold is different, and systematic trends in halo shape with mass introduce a net mass 
dependence in the collapse threshold, which can be fitted by: 
$$\delta_c(M,z) = \sqrt{a}\delta_{sc}[1 + b(a\nu^2)^{-\alpha}],$$ 
with $a \sim 0.707$, $b = 0.485$ and $\alpha \sim 0.615$, where $\delta_{sc}$ is the normal spherical collapse threshold \cite{Sheth01, Shethbarrier}. 
Now the change in barrier height $\Delta(\delta_c)  = \delta_c(M_1,z_1) - \delta_c(M_2,z_2)$ 
will depend on both masses and both redshifts separately, complicating the merger tree calculations.
We can simplify the calculations, however, by taking $\alpha = 0.5$ and using an approximation 
to the change in collapse threshold: $$\Delta(\delta_c) \sim  \sqrt{a}\Delta(\delta_{sc}(z)) + b\Delta\sigma.$$
For a fixed jump in redshift $z_1 \rightarrow z_2$, finding the exact solution to this equation still requires iteration since $\Delta\sigma$ itself depends on $\Delta\delta_c$, but using the spherical value as an approximation to $\Delta\sigma$, the approximate change in threshold can be calculated in one pass. Comparison with simulations [Taylor, in preparation] shows that this approximate ellipsoidal barrier provides a good match to measured halo merger probabilities at high values of $\Delta\nu$, while the spherical barrier works well at low values, so in practice the code used here switches between the two forms at $\Delta\nu \sim 0.15$. Using these adjustments to the merger probability, the correct mass dependence can be included in the merger trees, improving their age distribution and MAHs. 

\begin{figure*}
\begin{center}
\includegraphics[width=0.485\linewidth]{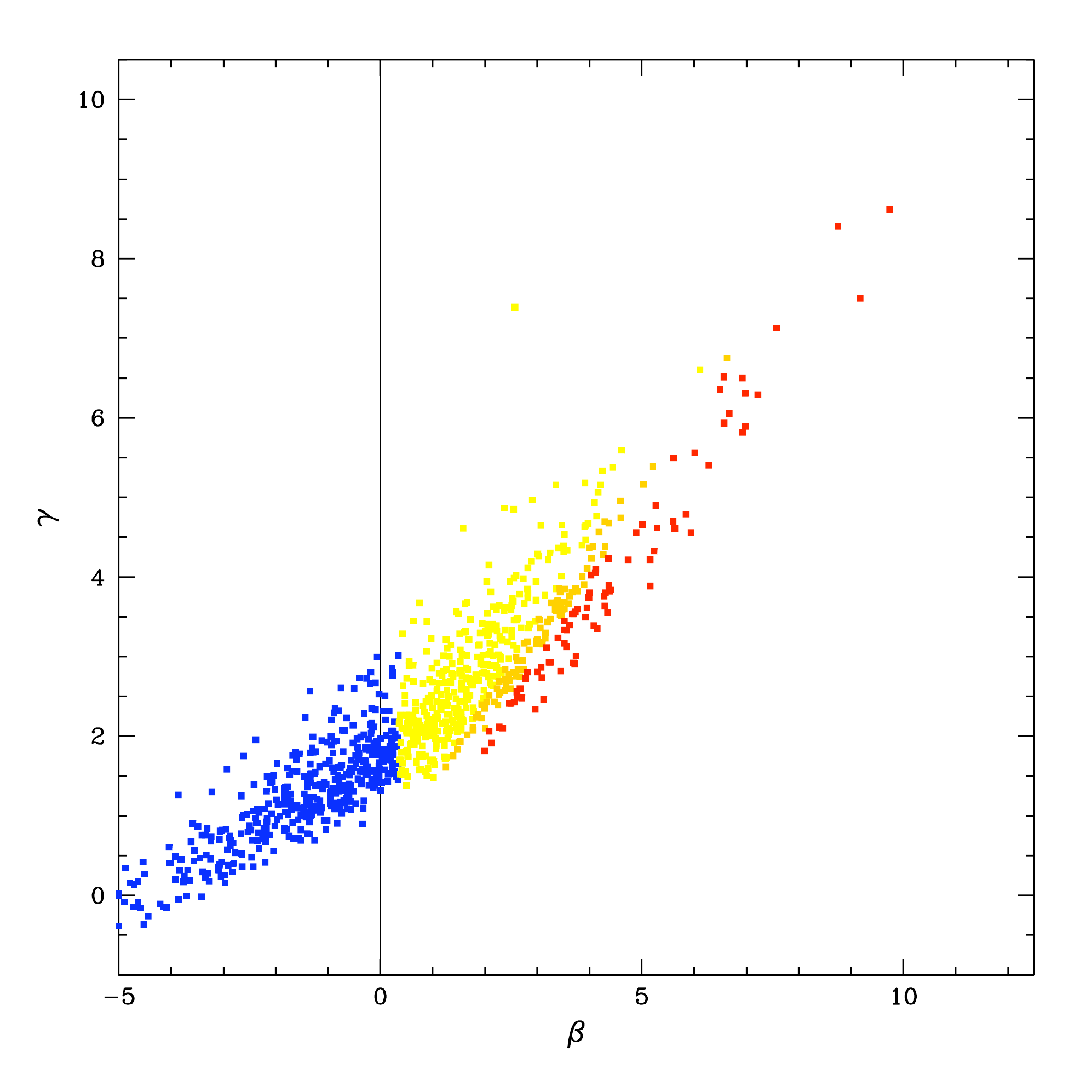}
\includegraphics[width=0.485\linewidth]{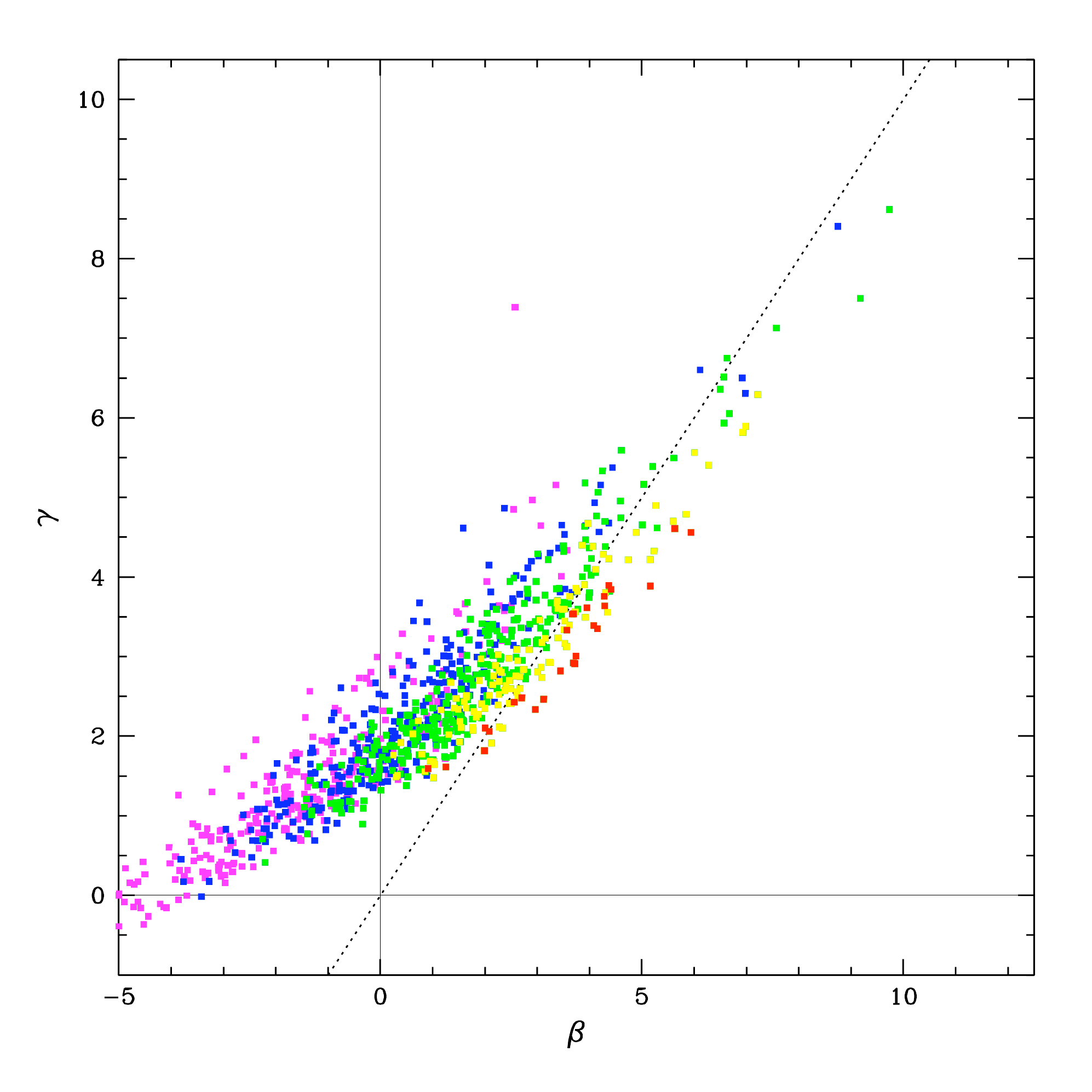}
\caption{Distribution of parameters describing the mass accretion histories of 1000 randomly generated merger trees. (Left) Colors correspond to the classes defined by Mc Bride et al.~ in \cite{Mcbride}: blue -- class I, yellow -- class II, orange -- class III, red -- class IV. (Right) colors correspond to the formation redshift $z_{50}$: magenta -- $z_{50} < 0.3$, blue -- $z_{50}$ = 0.3--0.5, green -- $z_{50}$ = 0.5--0.8, yellow-- $z_{50}$ = 0.8--1.1, red -- $z_{50} > $ 1.1. The dotted line indicates the locus of trajectories with $\gamma - \beta = 0$. For $\beta \gtrsim 1$, formation redshift is roughly constant along this lines parallel to this one.}
\label{fig:gammabeta}
\end{center}
\end{figure*}

Figure \ref{fig:gammabeta} shows the distribution of $\beta$ and $\gamma$ values fitted to the semi-analytic merger trees using $\chi^2$-fitting of their MAH. The distribution is very similar to the one found by McBride et al.~for halos in the Millenium simulation \cite{Mcbride}, although with a slight shift to higher values of $\gamma$ and/or lower values of $\beta$.  This may indicate that the semi-analytic trees are still slightly too young\footnote{Note that the distribution plotted in figure A1 of McBride et al.~includes halos of much lower mass, so this explains some of the difference.}. Mcbride et al.~classify MAH shapes into 4 classes, as indicated by the colors in the left-hand panel. Comparing to their statistics, 47\%, 36\%, 10\% and 8\% of the semi-analytic trees are in classes I, II, III and IV respectively, versus 18\%, 57\%, 17\% and 8\% of the most massive halos of the Millenium simulation. Thus, while the two distributions overlap significantly, the difference between them suggests that absolute numbers or age distributions derived from semi-analytic merger trees will need to be calibrated using simulated halos. The merger trees should capture relative trends in age and substructure with mass, redshift and cosmology, however.

It is also worth noting one other complicating factor in the $\beta$--$\gamma$ fit to MAHs. Both in the Millenium runs and in the semi-analytic models, measured parameter values cluster along a sequence roughly defined at high values of $\beta$ by $\beta = \gamma$ (dotted line in the right panel of figure \ref{fig:gammabeta}). Different points along lines parallel to this sequence actually represent almost indistinguishable fits to the numerical MAH, so it is not clear that $\beta$ and $\gamma$ represent the best parameterization of an accretion history. The difference $\beta-\gamma$ correlates much more closely with physical quantities like the formation redshift $z_{50}$ by which a halo had assembled 50\% of its present-day ($z=0$) mass, as shown by the colours in the right-hand panel.

\subsubsection{Subhalo Evolution}

Given a sequence of mergers from a merger tree, the next question is, how do these affect halo structure? This can be divided in two distinct questions: how do the overall properties (mass, concentration, shape, etc.) of a halo change during a merger, and how much dense substructure survives the merger? 

The model introduced in \cite{TB04} assumes the halo mass (really the gravitational potential) 
increases instantaneously at the moment of the merger and ignores changes in shape, 
treating the halo potential as spherical. For concentration, it uses a running fit of the MAH, 
together with the concentration model of \cite{Wechsler} (chosen for its simplicity) at each redshift. 
These choices give only a crude approximation to full 3-dimensional behavior of halos seen in 
simulations, but they do capture some of its essential features.

The evolution of merging subhalos is complex, and is discussed in detail in \cite{TB01, TB04}. Briefly, halo orbits are calculated in the (evolving) potential of the main system; dynamical friction gradually reduces the energy and angular momentum of subhalos, particularly massive ones; tidal stripping removes mass from the outer parts of satellites; tidal heating modifies their internal structure and accelerates mass-loss; and finally encounters and collisions also contribute to mass loss and 
scatter subhalo velocites. This combination of physics is required to produce realistic subhalo distributions that match high-resolution simulations \cite{TB05a, TB05b}. 

\subsection{The Subhalo Mass Function and Radial Distribution}

Halos retain traces of successive mergers, both in the positions and in the velocites of dark matter particles. Velocity substructure, in the form of coherent streams of particles on correlated orbits, is expected to be common but is probably only detectable in the halo of the Milky Way, where it can modify count rates in direct detection experiments (e.g.~\cite{Morgan, Helmispring, Zemp, Vogelsberger}). Real-space substructure consists of the dense cores of subhalos surviving from previous mergers. Since CDM structure formation proceeds from smaller mass halos to larger mass halos as the universe evolves from high density to low density, massive low-density halos at low redshift will contain the undigested cores of many low-mass, high-density halos, which are stable against tidal stripping provided they have a few times the mean density of the background (e.g.~\cite{Hayashi}). 

\begin{figure*}
\begin{center}
\includegraphics[width=0.485\linewidth]{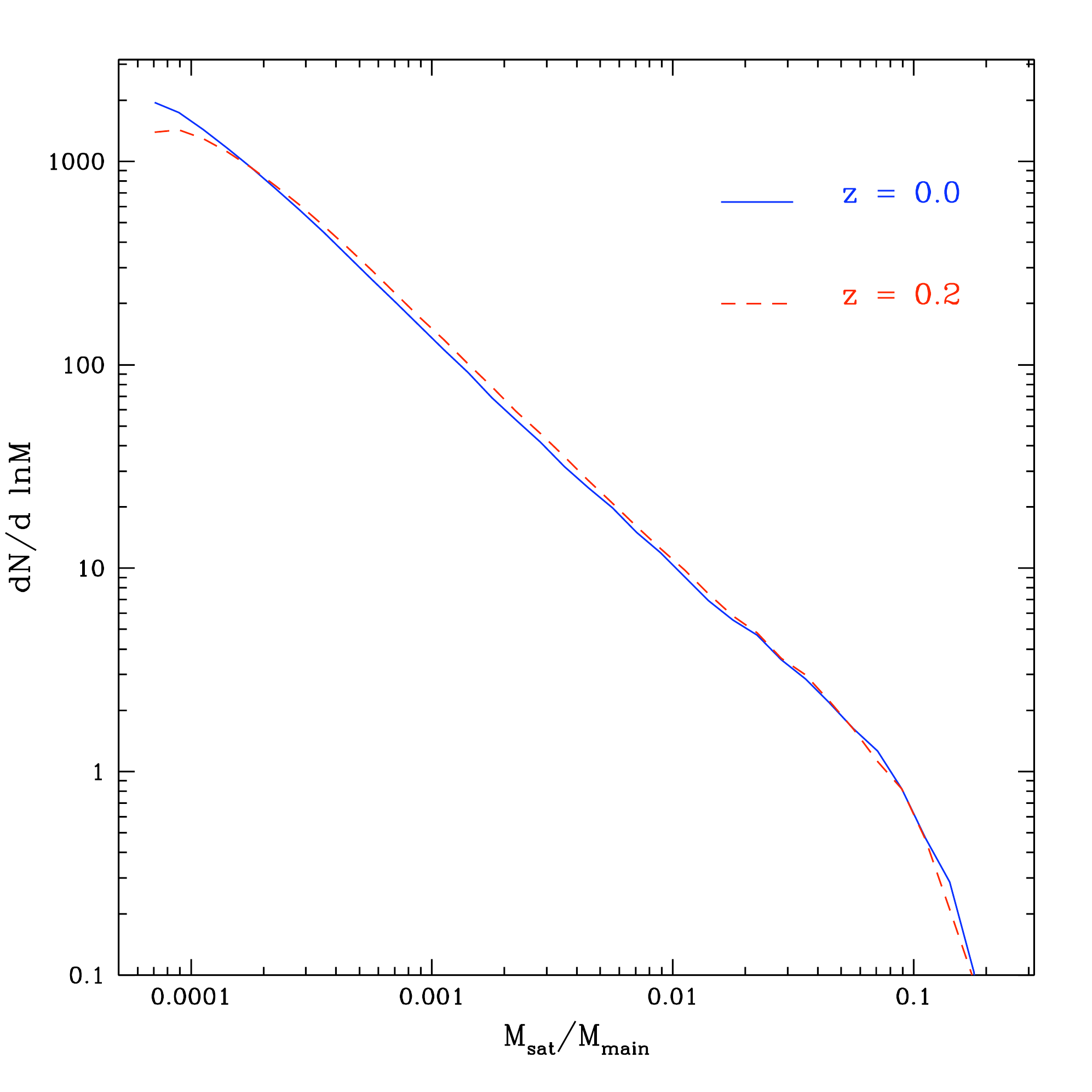}
\includegraphics[width=0.485\linewidth]{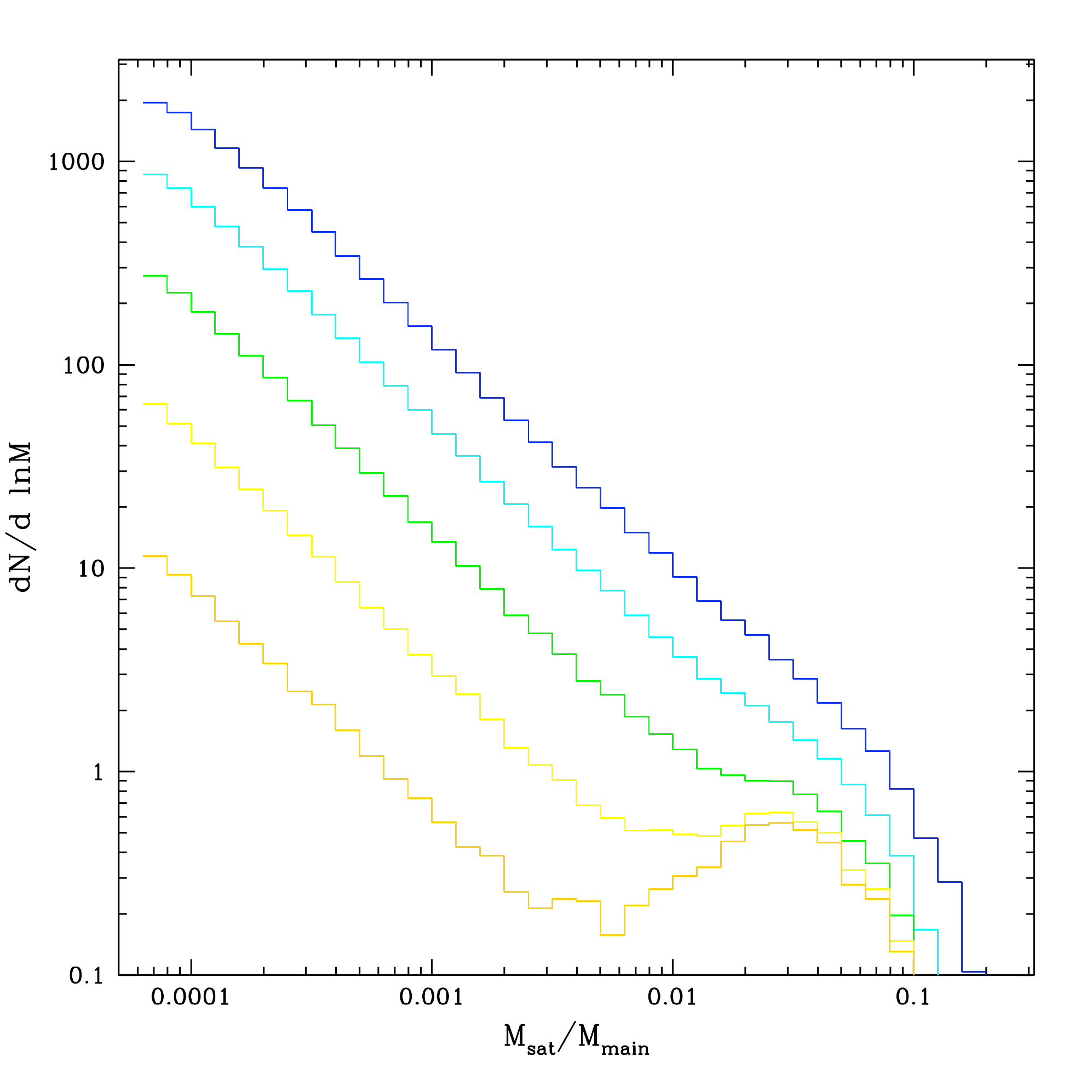}
\caption{(Left) The average differential mass function of subhalos within the virial radius for halos at redshift $z = 0$ (solid blue line) and $z = 0.2$ (dashed red line). (Right) The average differential mass function for halos at radii less than 1.0, 0.5, 0.25, 0.125 and 0.0625 times the virial radius (different histograms, from top to bottom). The excess of massive systems at small radii is clearly visible.}
\label{fig:Nm}
\end{center}
\end{figure*}

The mass spectrum and spatial distribution of these cores, or subhalos, are particularly simple, as illustrated in figure \ref{fig:Nm} which shows average mass functions for an ensemble of 3000 semi-analytic halo models. Relative to the mass of the main halo, the distribution of subhalo masses is a power-law with an exponential cutoff around $M_{sat}/M_{main} \sim 0.1 $. Normalized in this way, the mass function is almost invariant with redshift and varies only slowly with halo mass (see also \cite{TB05a, TB05b, Gao04, Giocoli, Gaorecent}). The spatial distribution with halo-centric radius (right-hand panel) is also very simple until one reaches 10--20\% of the virial radius from the centre. A small population of massive subhalos ($\sim 1$ or less per halo on average, with masses greater than 1\% of the main halo mass) is predicted in this region. These are systems which have been dragged in by dynamical friction and are in the process of disruption. The semi-analytic models ignore the effects of baryons, however, which should be strong in the central region. Thus it is not clear how many of the central merging systems exist in real cluster or galaxy halos -- hydrodynamic simulations would help clarify the situation here.

\subsection{Evolution with Time}

\begin{figure}
\begin{center}
\includegraphics[width=0.5\linewidth]{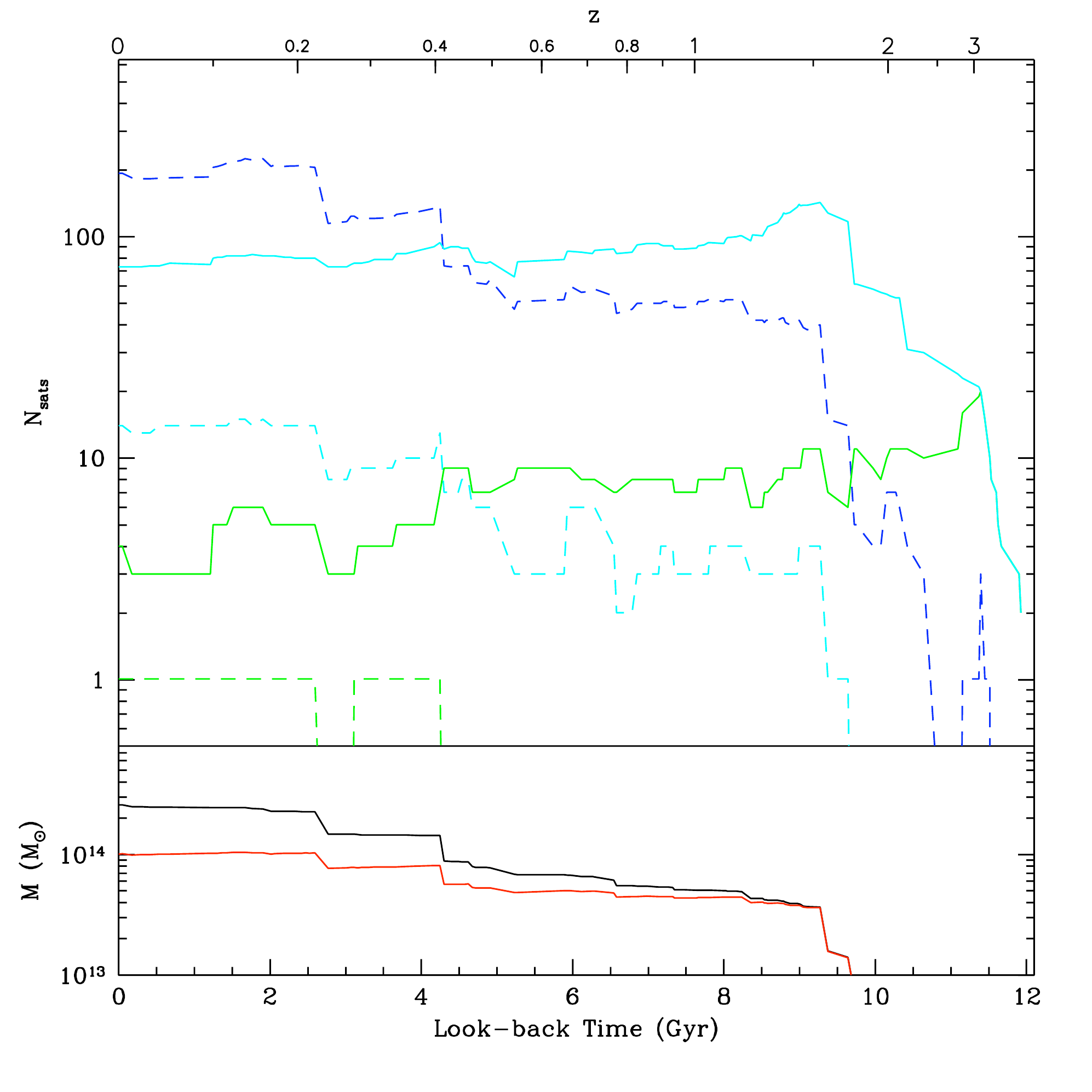}
\caption{The evolution of a single cluster halo versus time. The bottom panel shows how the total mass (upper black line) and the mass within the central 390 kpc (lower red line) build up with time. The upper panel shows how the number of satellites changes with time. The dashed curves are for satellites with masses in excess of 
$10^{11} M_\odot$, $10^{12} M_\odot$ and $10^{13} M_\odot$ respectively from top to bottom. The solid curves are for satellites with masses of more than 0.1\% or 1\% of the main system's mass respectively, from top to bottom.}
\label{fig:evolution}
\end{center}
\end{figure}

The subhalo mass function is not completely static, but changes systematically with time. While 
low-mass halos reach an equilibrium in mass, energy and angular momentum loss after a few 
orbits in the main halo, massive subhalos evolve quickly due to dynamical friction. As a result, 
massive substructure is a key indicator of recent growth and merging. 
Figure \ref{fig:evolution} shows the evolution of a single semi-analytic model halo over time. 
The lower panel
shows how the total mass of the main halo (upper, black line) and the mass within the central 390 kpc build up with time. The upper panel shows how the number of satellites changes with time. The dashed curves are for satellites with masses in excess of $10^{11} M_\odot$, $10^{12} M_\odot$ and $10^{13} M_\odot$ respectively, from top to bottom. The solid curves are for satellites with masses of more than 0.1\% or 1\% of the main system's mass respectively, from top to bottom. As the halo grows, its total number of satellites increases, but the mass ratio of each individual subhalo decreases, both because it loses mass through tidal stripping and because its parent halo increases in mass. Thus the amplitude of the halo mass function expressed in terms of the ratio $M_{sat}/M_{main}$ decays with time, particularly for large values of $M_{sat}/M_{main}$  (e.g.~the green solid curve in the upper panel).

\begin{figure}
\begin{center}
\includegraphics[width=0.5\linewidth]{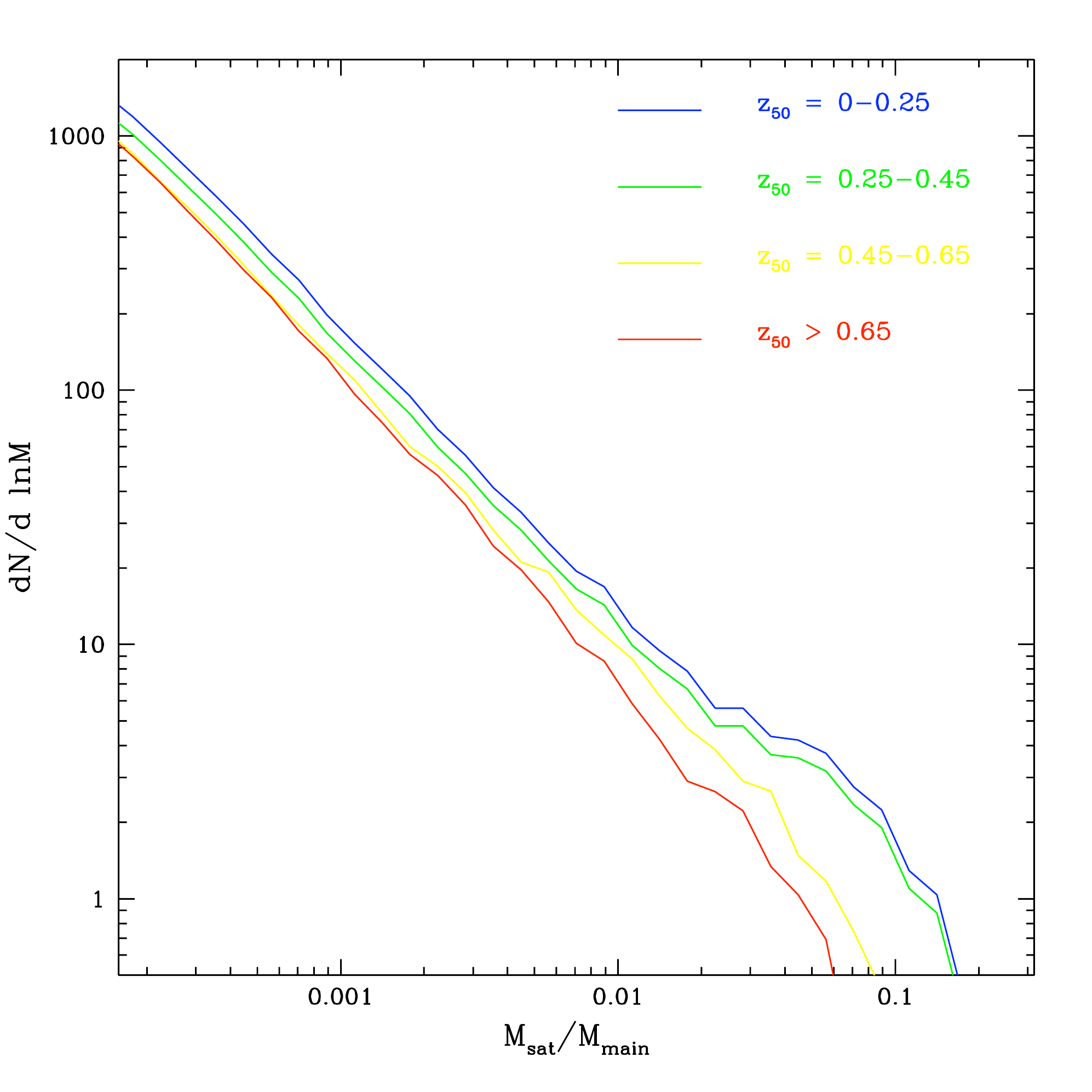}
\caption{The differential mass function of subhalos for systems with different ranges of formation epoch: $z_{50} = $0--0.25 (blue), $z_{50} = $0.25--0.45 (green), $z_{50} = $0.45--0.65 (yellow) and $z_{50} > $0.65 (red).}
\label{fig:Nm_age}
\end{center}
\end{figure}

As a result of these trends, the slope of the subhalo mass function changes with time. This suggests a first observational test of halo age, based on substructure. If one can measure subhalo masses for individual halos, particularly at the high mass end, then the amplitude of the mass function will correlate with evolutionary history. The correlation is strongest with recent tracers of halo growth such as $z_{90}$, the redshift by which a halo built up 90\% of its present-day ($z=0$) mass \cite{TB05a}, but is evident even for longer-term tracers like $z_{50}$, as shown in figure \ref{fig:Nm_age}. (Note that this is a differential mass function; integrating the contribution from all halos would produce an even stronger signal.) I explore this idea further below.

\subsection{Inferring Age from Observation of Substructure}

\begin{figure*}
\begin{center}
\includegraphics[width=0.50\linewidth]{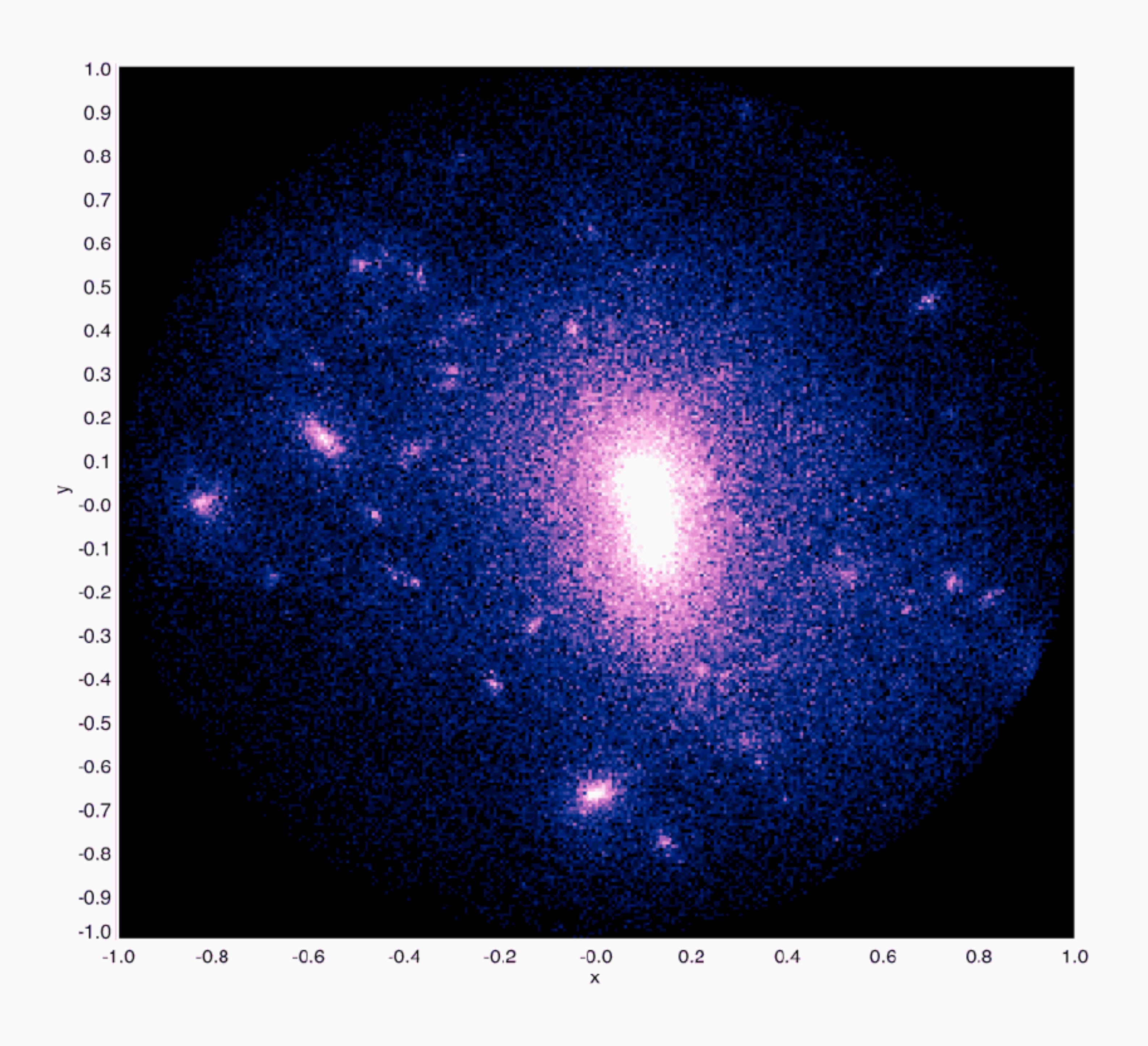}
\includegraphics[width=0.47\linewidth]{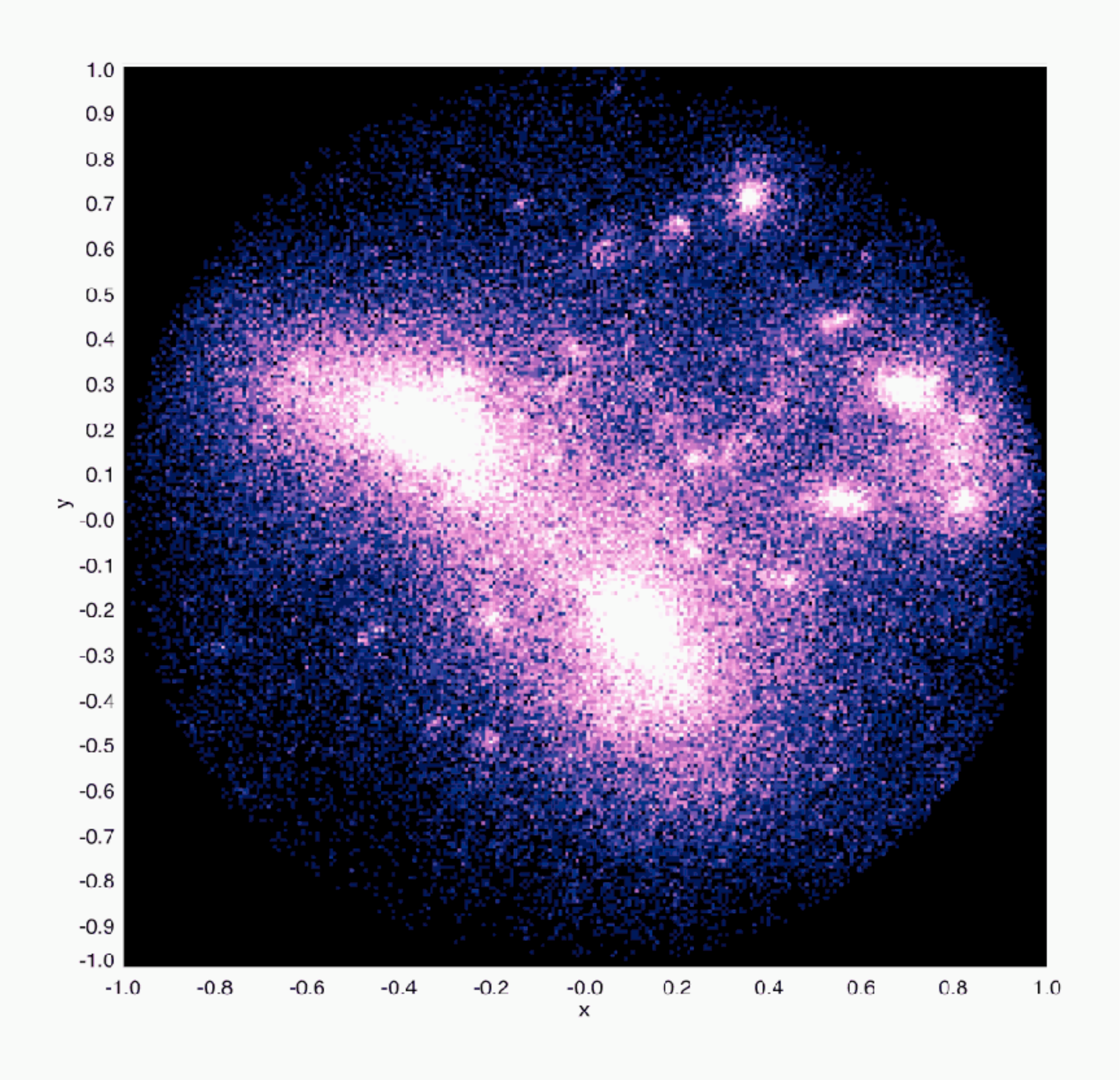}
\caption{Two cluster halos extracted from an n-body simulation at $z = 0$. One is isolated and relaxed (left), while the other is in the process of forming through a major merger (right). In each case, only particles within 1 Mpc of the halo are plotted.}
\label{fig:simulations}
\end{center}
\end{figure*}

The differences shown in figure \ref{fig:Nm_age} suggest there should be a strong correlation between the {\it total} amount of substructure in a halo and its formation history. Numerical studies find similar results, albeit with large halo-to-halo scatter \cite{Gaorecent, Ishiyama}. At a broader level, it has long been appreciated that cluster substructure indicates recent growth \citep{oldcluster}. Figure \ref{fig:simulations} shows two regions extracted from a recent n-body simulation run with WMAP-7 parameters. For clarity, only the particles within 1 Mpc of the centre of each region are plotted. The left-hand region contains relatively isolated halo, while the right-hand region catches an ongoing merger between two components. Even with an imperfect reconstruction of the matter density, e.g. via lensing 
measurements of the convergence field or X-ray measurements of the gaseous emission in each region, observations could clearly tell one from the other. The challenge is to detect more subtle signs of recent growth, and to do so for a large sample of halos. 

Galaxy clusters are a tempting target for this sort of study, for several reasons. Luminous baryons can be used to trace at least part of the substructure in cluster halos down to very small values of $M_{sat}/M_{main}$, since individual galaxies have mass ratios of $10^{-5}$ or less relative to the cluster. Hot gas provides an independent measure of the shape of the main halo potential, as does lensing. Lensing is the more appealing of these tracers, since it avoids possible offsets or biases between the baryonic matter distribution and the dark matter distribution, but it requires deep observations of massive clusters at moderate redshifts. Many groups have used lensing to reconstruct the matter distributions in famous systems like Abell 1689 (e.g.~\cite{Broadhurst, Diego}) or the Bullet Cluster \cite{Clowe}. The LoCuSS survey (http://www.sr.bham.ac.uk/locuss) is a more general attempt to measure the shape of the potential in a large sample of cluster halos and to relate the mass distribution in clusters to their dynamical state and
galaxy populations.

These studies combine large scale weak-lensing measurements (measurements of weak distortions in galaxy shape only apparent in averages over hundreds of galaxies), to get the overall mass distribution 
in the cluster, with strong lensing observations, which can reveal small-scale features in the central part of the cluster. The relative contribution of structure to the projected mass distribution in the cluster can 
be estimated by comparing a smooth ellipsoidal model  to the full mass distribution inferred from the observations. For a particular choice of aperture (roughly 390 kpc in the case of LoCuSS, based on the field of view of the camera and the redshifts of the clusters), the difference between the two, normalized to the total mass, gives the substructure fraction $f_{sub}$. It is straightforward to calculate the 
distribution of 
$f_{sub}$ in the semi-analytic models, assuming a perfect set of observations which reconstruct the full mass distribution to high precision over the entire aperture. It is less clear (and the subject of ongoing work) how gaps in the lensing map affect measurements of $f_{sub}$ in real 
clusters. Nonetheless, preliminary comparisons of the models with the data show good general agreement \cite{ST08}.

\begin{figure*}
\begin{center}
\includegraphics[width=0.485\linewidth]{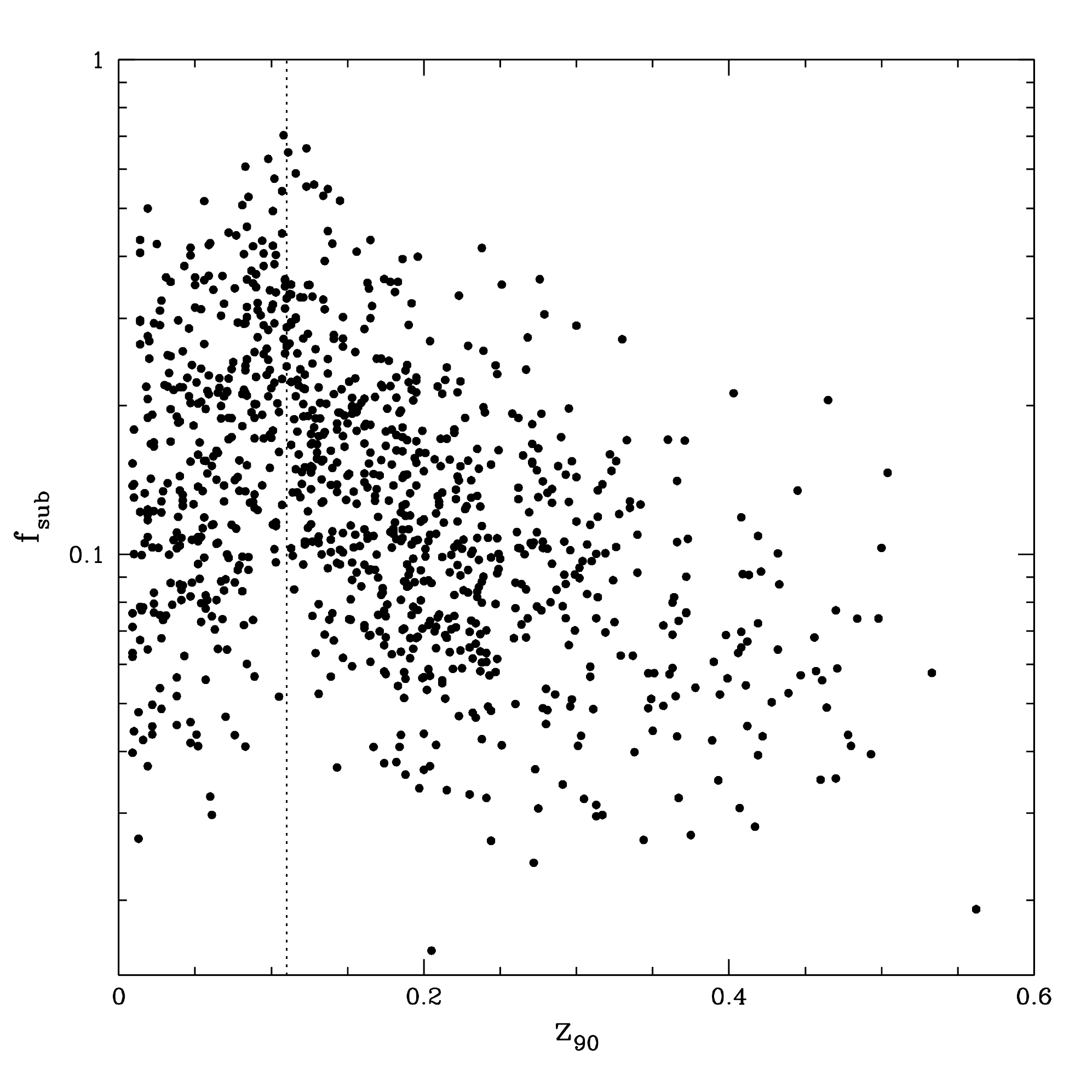}
\includegraphics[width=0.485\linewidth]{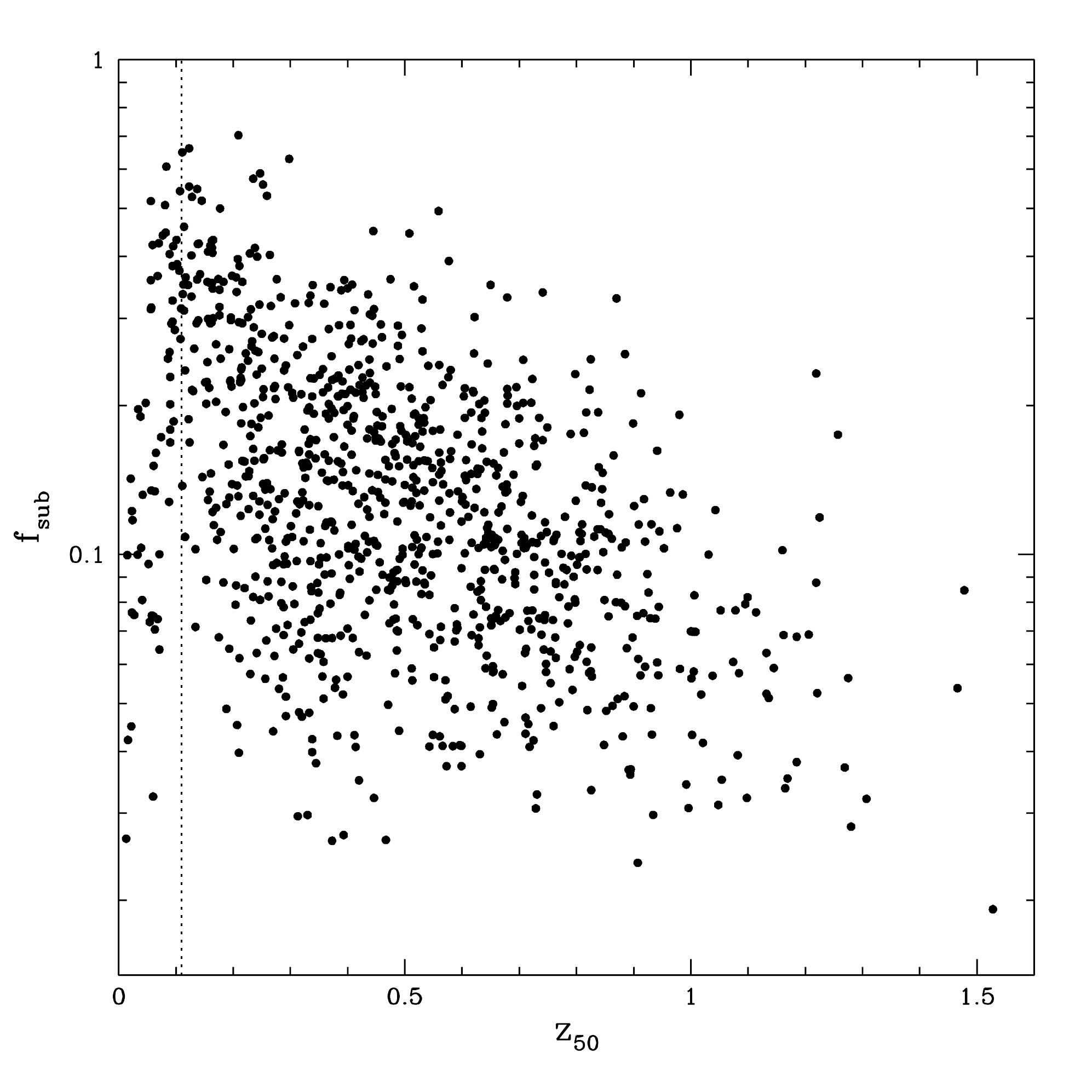}
\caption{Lensing substructure fraction versus the formation reshifts $z_{90}$ (left) and $z_{50}$ (right). The dotted line indicates the epoch approximately one infall time before present.}
\label{fig:fsub}
\end{center}
\end{figure*}

The predicted distribution of $f_{sub}$ is roughly log-normal, as discussed below. The substructure fraction correlates strongly with the formation redshifts $z_{90}$ and $z_{50}$ defined previously, although in both cases the relationship is not quite monotonic (figure \ref{fig:fsub}). As discussed in \cite{ST08}, for systems that have formed very recently, merging material will not have reached the centre of the halo yet. Thus systems with $z_{90} < 0.1$ sometimes have relatively low central substructure fractions. The dotted lines in the two panels indicate the formation redshift  such that merging 
substructure has reached the centre of the halo and is on its first pericentric passage. Beyond this point the correlation between formation epoch and substructure fraction is clear and monotonic. Only the 
vertical axis of these figures corresponds to an observable quantity, $f_{sub}$, but clearly by 
binning in $f_{sub}$ one can select samples of clusters with very different age distributions. Thus 
measurements of $f_{sub}$ represent one realistic approach to determining the underlying distribution of halo ages.

Finally, I note that similar work constraining substructure fractions with lensing observations may also be possible on much smaller mass scales, in the halos of individual galaxies. While individual galaxies are not massive enough to be detected through weak lensing, 
their central regions reach surface densities high enough to produce strong lensing of background galaxies or quasars. Substructure in the halo of a lensing system can produce variations in brightness (e.g.~\cite{DalalKochanek, Chiba, Metcalf}), image position (e.g.~\cite{Chen}), or the time delays between multiple images of variable background sources (e.g.~\cite{Keeton}) -- see \cite{Lensrev} for a recent review. Separating the signals from dark and luminous substructure is more complicated on this scale, since there is possible contamination from microlensing by stars in the lensing galaxy, and since many subhalos large enough to produce detectable variations in the lensing potential do not host galaxies bright enough to detect at cosmological redshifts. Nonetheless, with forthcoming large samples of strong lenses, this may be promising avenue for measuring dark substructure on very small scales (e.g.~\cite{Vegetti}).

\section{Prospects}
\label{sec:discuss}

The simple, universal distributions of halo parameters -- concentration, shape, spin, substructure -- that emerge from numerical simulations, and the strong correlations found between these quantities, suggest the one or two main properties determine much of the internal structure of dark matter halos. 
Section \ref{sec:universal} reviewed the evidence that a halo's concentration parameter is determined directly by its mass accretion history, and in section \ref{sec:obs} I shown how substructure should also reflect formation history. The semi-analytic models used to study this relationship are somewhat approximate and need confirmation from a large suit of numerical simulations [Wong et al.~in preparation], 
but the predicted trends are clear and match those seen in earlier numerical work. Thus in these two examples, the fundamental parameter which accounts for most of the scatter in halo properties seems to be age, in the broad sense encapsulated by the mass accretion history.

Mass accretion histories can in principle contain an arbitrary number of independent degrees of freedom. Based on the numerical fits discussed in section \ref{sec:universal}, real examples seem to form a much more limited set with only one (or possibly two) main degree(s) of freedom. Likewise, while the shape of the substructure mass function can in principle have many degrees of freedom, it seems to follow a single-parameter sequence in practice, and of course scalar properties like concentration necessarily form 1-dimensional distributions. All this suggests there may be a single `best' measure of halo age, `best' in the sense that it captures most of the variety in halo properties. This exact identity of this parameter, its physical interpretation and the best observational estimators for it remain to be determined.

Assuming a single-parameter sequence of halo `age' or degree of relaxation exists, why bother 
measuring it? Understanding how halos achieve their particular internal states is intrinsically interesting, 
of course, but information about halo structure and dynamics also has several immediate practical applications.
As discussed in section \ref{sec:nonlinear}, halo occupation (HOD) models predict or interpret galaxy 
clustering by matching galaxies to halos, and this approach is also useful in producing `mock' 
catalogues of galaxies from dark-matter-only simulations. These models normally assume that
mass is the most important halo property to match, despite evidence that halo age is correlated with large-scale clustering and environment \cite{Gao05, Maulbetsch, Jing07, Li08, Dalal08, Wechsler08, Faltenbacher}. Better definitions and metrics for age, together with simulations of galaxy formation in halos with different formation histories, would clarify the connection between the age of the host halo and its resident galaxy's properties.

\begin{figure}
\begin{center}
\includegraphics[width=0.485\linewidth]{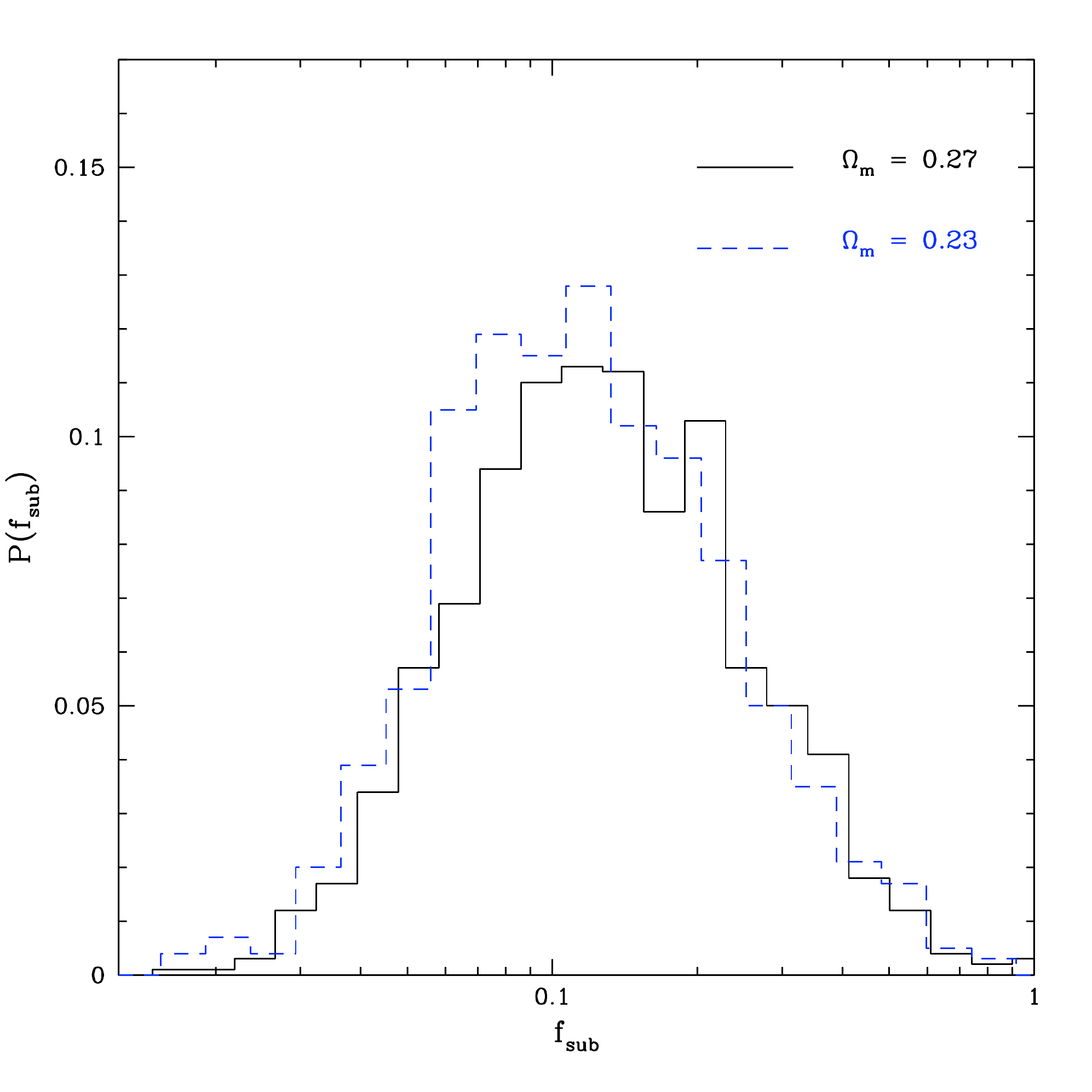}
\includegraphics[width=0.485\linewidth]{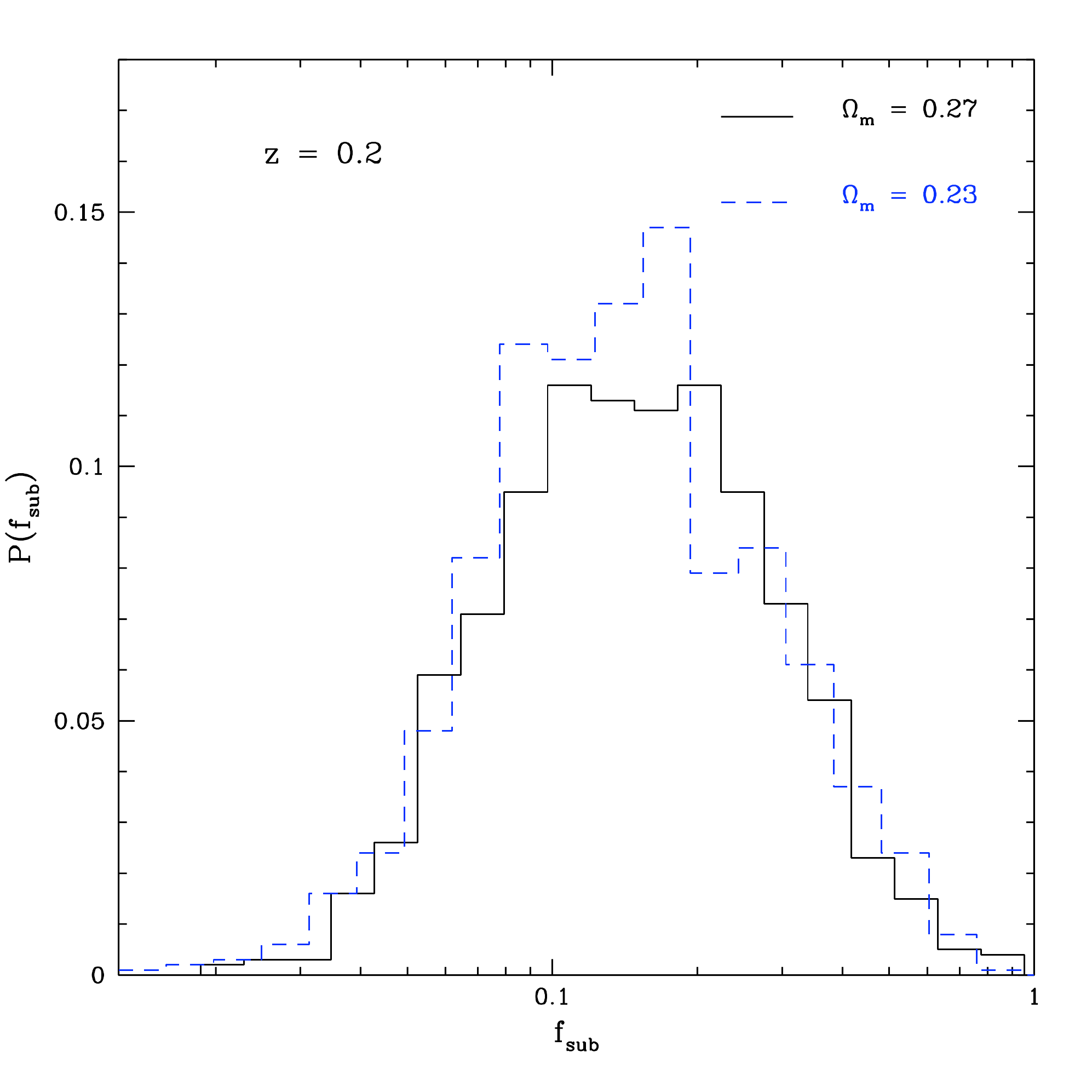}
\caption{Distributions of lensing substructure fraction at $z = 0$ (left) and $z = 0.2$ (right). 
The solid histogram is for 1000 merger trees in a cosmology with $\Omega_m = 0.27, \sigma_8 = 0.80$. The dashed histogram is for 1000 merger trees in a cosmology with $\Omega_m = 0.23, \sigma_8 = 0.85$. These two cosmologies are chosen to produce the same abundance of massive clusters at low redshift, so cluster number counts alone would not distinguish between them.}
\label{fig:fsubcosmo}
\end{center}
\end{figure}

A second practical application is testing standard structure formation and breaking parameter degeneracies. One simple example is the age distribution of galaxy clusters. Clusters form through the amplification of peaks in the early density field by the growth factor $D(z)$. Measuring the abundance of clusters constrains the product of the power spectrum and the growth factor, leading to a degeneracy 
between $\Omega_m$ and $\sigma_8$\footnote{$\sigma_8 = \sigma(M(R))$ for $R = 8\,h^{-1}$Mpc, where $h = (H_0/100$ km s$^{-1}$ Mpc$^{-1}$).} (or for more general cosmologies, between late-time growth and initial power). At fixed cluster abundance, there is a physical difference between cosmologies with low $\Omega_m$ and high $\sigma_8$ and those with high 
$\Omega_m$ and low $\sigma_8$, however; in the former case clusters form earlier, and should be more relaxed, rounder, more concentrated and have less substructure. Figure \ref{fig:fsubcosmo} shows how the substructure fraction distributions reflect the difference in cluster ages for two very similar cosmologies located along the degeneracy in the $\Omega_m$--$\sigma_8$ plane. The shift in the distribution is subtle (partly because substructure is being measured in projection -- see section \ref{sec:obs}), but should be measurable with a sample of {\it O}(100) clusters, provided the effect can be calibrated carefully in simulations.

In the longer term, a more fundamental  goal of the study of non-linear structure is to understand dark matter itself. Non-linear structure -- or halo properties more specifically -- are connected to the fundamental physics of dark matter in several ways. Throughout this article I have assumed that all dark matter is a single particle of a `plain vanilla' sort, that is cold, collisionless and stable over cosmological timescales and on all the length scales we can probe. In fact a much greater range of possibilities exist -- see \cite{Bertone} for a comprehensive review. Real candidates could behave quite differently on small scales or at high densities, their properties could vary over cosmological time, or they could start off with a quite different initial distribution in the early universe. Detailed studies of dark matter halos will help constrain all three of these possibilities.
 
There are a number of particle properties which would only become evident at high densities or on small spatial scales. Supersymmetric dark matter particles such as neutralinos are their own anti-particle and can annihilate with one-another, producing gamma-rays and other secondary particles with energies in the GeV--TeV range. This emission is strongly weighted to the densest part of halos and thus depends sensitively on the amount of substructure within a halo (see e.g.~\cite{Kuhlenrev} for a recent review). In principle, dark matter particles could also have important collisional cross-sections with themselves and/or normal matter, or their coupling to gravity might even be different. All these possibilities are already quite strongly constrained by current observations, however. Observations by the Fermi Gamma-ray telescope constrain the annihilation cross-section for dark matter \cite{Fermi}, observations of the bullet cluster constrain the elastic scattering cross-section \cite{Clowe} and observations of the disruption of the Sagittarius dwarf galaxy constrain a modified gravitational coupling \cite{Kesden}. As our measurements of halo structure improve, probing denser substructure and smaller mass scales, these constraints will grow stronger. The time evolution of dark matter properties is slightly less well constrained, but since most of the evidence for dense substructure in dark matter halos comes from low redshift, this suggests that dark matter particles have remained stable over the age of the universe. Better observational constraints on dense halo substructure and the overall distribution of halo ages could strengthen this argument considerably.

The initial conditions for structure formation are a much more open question. Modifications to dark matter couplings, or variations in the equation of state of the universe at early times, could leave their imprint in the power spectrum of fluctuations from which halos formed. Warm dark matter, strongly annihilating or rapidly decaying dark matter, collisional dark matter and other variations on the plain vanilla model all produce a truncated power spectrum and a minimum scale to structure formation. This would eliminate halo substructure below that scale, and produce rounder and more relaxed halos on slightly larger scales. There have been many appeals to modified physics of this kind to explain apparent discrepancies between the distributions of dark and luminous matter in galaxy cores (as reviewed in \cite{Blok}) or in the halo of the Milky Way (as reviewed in \cite{Kravtsovrev}). These arguments have the problem that the baryonic tracers themselves are expected to disappear on small scales, due to strong negative feedback effects in star formation and galaxy formation on these scales. When we fail to observe dense, small-scale dark structure in the universe, it is unclear whether this is telling us something about the behavior of dark matter on these scales, or something  about the behavior of baryons. Current and future lensing experiments will be crucial for resolving this problem and establishing or disproving definitively the existence of dark structure on sub-galactic scales. This in turn should lead to strong constraints on 
warm or collisional dark matter, or similar models.

A generation of numerical experiments and analytic work have made remarkable progress in understanding non-linear structure formation and the `deeply' non-linear internal properties of dark matter halos. While the field lacks the analytical simplicity of linear theory and may seem less suited to precision cosmology as a result, some simple patterns have emerged from simulations. Dark matter halo properties correlate strongly with their formation history and may even form a single-parameter sequence in `age', provided this parameter can be defined clearly. With powerful new observational measures of halo structure and substructure becoming available, small-scale non-linear structure formation may be the next great source of tests of cosmology, dark matter and fundamental physics.

\section*{Acknowledgements}

This work was supported by a Discovery Grant from the Natural Sciences and Engineering Research Council of Canada. I thank Anson Wong for providing the data used to make figure \ref{fig:simulations}. I am also happy to acknowledge many useful conversations about dark matter models, astrophysical constraints on dark matter and the prospects for dark matter identification, with the participants of the Keck Institute for Space Studies workshop ``Shedding Light on the Nature of Dark Matter", which was funded by the W. M. Keck Foundation. Finally, I thank P.~Salucci for clarifying the original evidence for dark matter in rotation curves.

\bibliography{references}
\bibliographystyle{ieeetr}

\end{document}